Special feature reviews

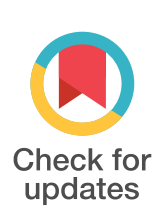

**Author for correspondence:**
Chen Li
e-mail: [redacted]

# Locomotor transitions in the potential energy landscape-dominated regime

Ratan Othayoth, Qihan Xuan, Yaqing Wang and Chen Li

Department of Mechanical Engineering, Johns Hopkins University, 3400 N. Charles Street, Baltimore, MD 21218, USA

RO, 0000-0001-5431-9007; QX, 0000-0002-1075-1516; YW, 0000-0001-6884-323X; CL, 0000-0001-7516-3646

To traverse complex three-dimensional terrain with large obstacles, animals and robots must transition across different modes. However, most mechanistic understanding of terrestrial locomotion concerns how to generate and stabilize near-steady-state, single-mode locomotion (e.g. walk, run). We know little about how to use physical interaction to make robust locomotor transitions. Here, we review our progress towards filling this gap by discovering terradynamic principles of multi-legged locomotor transitions, using simplified model systems representing distinct challenges in complex three-dimensional terrain. Remarkably, general physical principles emerge across diverse model systems, by modelling locomotor–terrain interaction using a potential energy landscape approach. The animal and robots' stereotyped locomotor modes are constrained by physical interaction. Locomotor transitions are stochastic, destabilizing, barrier-crossing transitions on the landscape. They can be induced by feed-forward self-propulsion and are facilitated by feedback-controlled active adjustment. General physical principles and strategies from our systematic studies already advanced robot performance in simple model systems. Efforts remain to better understand the intelligence aspect of locomotor transitions and how to compose larger-scale potential energy landscapes of complex three-dimensional terrains from simple landscapes of abstracted challenges. This will elucidate how the neuromechanical control system mediates physical interaction to generate multi-pathway locomotor transitions and lead to advancements in biology, physics, robotics and dynamical systems theory.

## 1. Introduction

To move about, animals can use many modes of locomotion (e.g. walk, run, crawl, slither, burrow, climb, jump, fly and swim) [1,2] and often transition between them [3,4]. Despite this multi-modality, the most mechanistic understanding of terrestrial locomotion has been on how animals generate [5–8] and stabilize [9–11] steady-state, limit cycle-like locomotion using a single mode.

Previous studies began to reveal how terrestrial animals stochastically transition across locomotor modes in complex environments. Locomotor transitions, like other animal behaviour, emerge from multi-scale interactions of the animal and environment across the neural, postural, navigational and ecological levels [12–14]. At the neural level, terrestrial animals use central pattern generators [15] and sensory information [16–18] to switch locomotor modes to traverse different media or overcome obstacles. At the ecological level, animals foraging across natural landscapes switch locomotor modes to minimize metabolic cost [19]. At the intermediate level, terrestrial animals transition between walking and running to save energy [20]. However, there remains a knowledge gap in how locomotor transitions in complex three-dimensional terrain emerge from physical interaction (i.e. terradynamics [21]) of an animal's body and appendages with the environment mediated by the nervous system. We lack theoretical concepts for thinking about how to generate and control locomotor transitions on the same level of limit cycles for steady-state, single-mode locomotion [22].

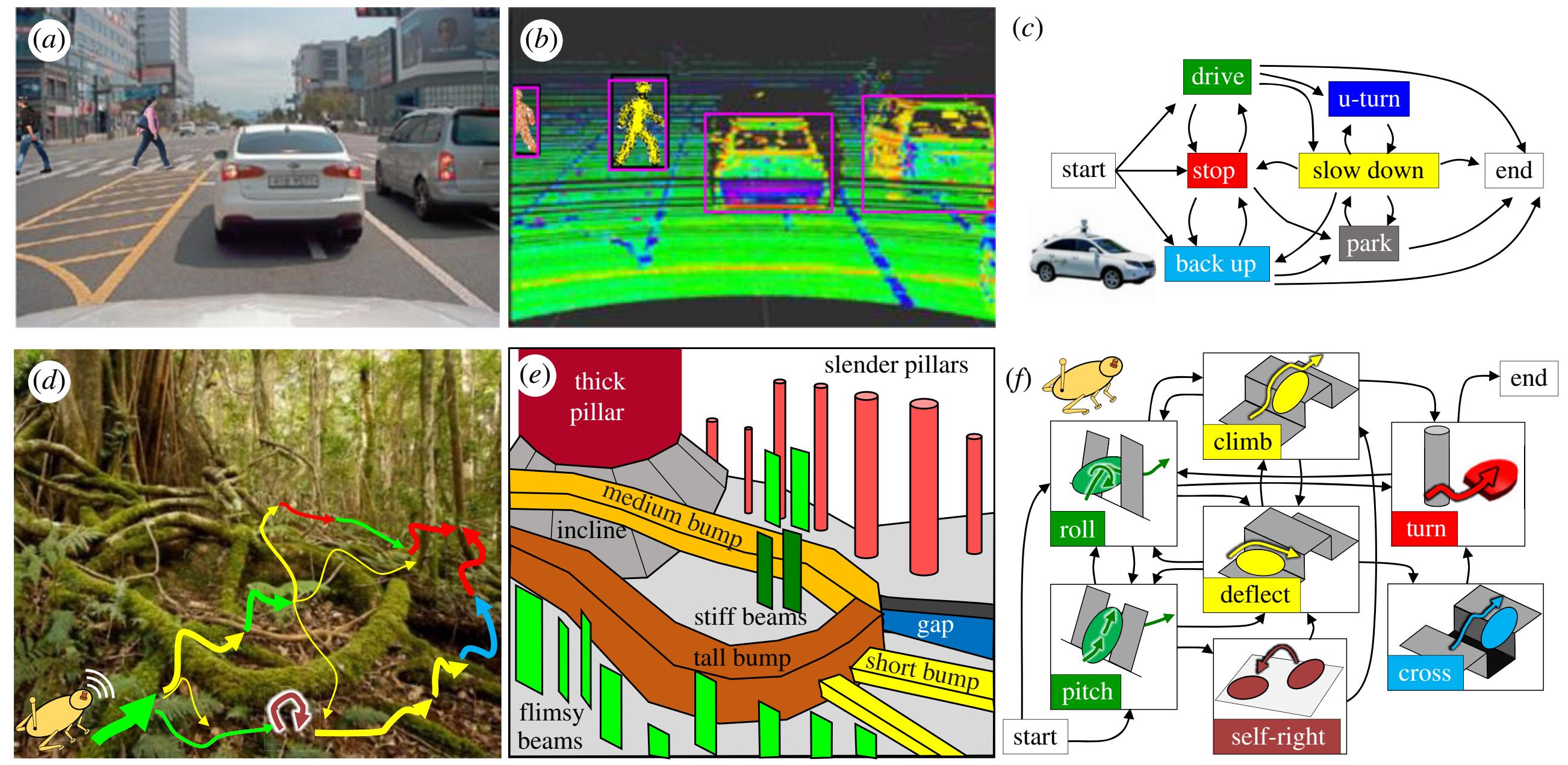

**Figure 1.** Multi-pathway transitions to avoid and traverse obstacles. (*a*) View from a self-driving car. (*b*) Geometric map scanned. (*c*) Multi-pathway driving transitions to avoid obstacles. (*d*) Envisioned capability of robot traversing complex three-dimensional terrain with many obstacles as large as itself. (*e*) Abstracted challenges from diverse large obstacles. (*f*) Envisioned multi-pathway locomotor transitions. Image credits: (*a*,*b*), Modified with permission from [23] under Creative Commons CC-BY license. (*d*) Modified with permission from Luke Casey Photography. (Online version in colour.)

Understanding of how to use physical interaction with complex three-dimensional terrain to generate and control locomotor transitions is also critical to advancing mobile robotics. Similar to personal computers in the 1970s, mobile robots are on the verge of becoming a major part of society. Wheeled robots like robotic vacuums and self-driving cars (figure 1*a*) already excel at avoiding sparse obstacles to navigate flat homes, streets and even unpaved roads, by scanning a geometric map of the environment (figure 1*b*) and acting upon it to transition between driving modes (figure 1*c*) [24]. This owes to the well-understood wheel–ground interaction physics [25,26]. Understanding of appropriate leg–ground physical interaction to generate and stabilize steady-state running and walking [5,6] enabled animal-like legged robot locomotion (such as from Boston Dynamics) on near-flat surfaces with small obstacles. However, despite progress in robot design, actuation and control for multi-modal locomotion [3], robots still struggle to make robust locomotor transitions to traverse obstacles as large as themselves, hindering important applications such as environmental monitoring in forests (figure 1*d*), search and rescue in rubble and extraterrestrial exploration through rocks. This is largely due to a poor understanding of physical interaction in complex three-dimensional terrain.

A physics-based approach by creating a new field of terradynamics [21] holds promise for filling this major gap. For aerial and aquatic locomotion of animals and robots, we understand fairly well their fluid–structure interaction thanks to well-established experimental, theoretical and computational tools, such as wind tunnel and water channel, aerofoil and hydrofoil, aero- and hydrodynamic theories, and computational fluid dynamics techniques [27]. By creating controlled granular media testbeds, robotic physical models [28,29], and theoretical and computational models, recent studies elucidated how animals (and how robots should) use physical interaction with granular media to move effectively both on and within sandy terrain (see [30] for a review). The general physical principles [30] and predictive physics models [21,30] not only advanced understanding of functional morphology [31–33], muscular control [34,35] and evolution [36] of animals, but also led to new design and control strategies [28,30,37–40] that enabled a diversity of robots to traverse granular environments.

Inspired by these successes, our group has been expanding the field of terradynamics to locomotion in complex three-dimensional terrain, by integrating biological experiments, robotic physical modelling and physics modelling (figure 2). Here, we review our approaches, progress and opportunities ahead. This review focuses on multi-legged locomotor transitions; for our work on limbless locomotion in three-dimensional terrain, see [42–47]. We studied the rainforest-dwelling discoid cockroach (figure 3*a*), which is exceptional at traversing complex three-dimensional terrain with diverse large obstacles such as vegetation, foliage, crevices and rocks [4]. Just like how understanding the aerodynamics of passive aerofoils provides a foundation for understanding flight control [60], we first focused on understanding passive mechanical interaction, which provides a foundation for understanding sensory feedback control (and the intelligence aspect of locomotor transitions in general). This is achieved by studying the animal in the rapid, bandwidth-limited escape [61] or emergency self-righting response and feed-forward-controlled robotic physical models. Although still at an early stage, our work begins to reveal general physical principles of locomotor transitions, which is remarkable considering that complex three-dimensional terrain is highly heterogeneous with diverse obstacles. Our work again demonstrates the power of interdisciplinary integration to discover terradynamic principles.

## 2. Experimental tools

### (a) Model terrain

To begin to understand complex physical interaction during locomotion in nature (figure 1*d*), we abstracted complex three-dimensional terrain as a composition of diverse large obstacles (figure 1*e*) that present distinct locomotor challenges. These include compliant beams [50,51], rigid pillars [52], gaps [53] and bumps [54]. To enable systematic experiments (as in a wind or water tunnel), for each model terrain, we created a testbed that allowed controlled, systematic variation of obstacle properties such as stiffness [50], geometry [52] and size [53,54] (figure 3*b*). In addition, because animals and robots often flip over when traversing large obstacles [4,52,55], we studied strenuous ground self-righting in which existing appendages must be co-opted [55–59]. Furthermore, we developed tools to address technical challenges in measuring locomotor transitions and locomotor–terrain interaction in complex three-dimensional terrain (figure 3*b*–*d*; electronic supplementary material, Text S1).

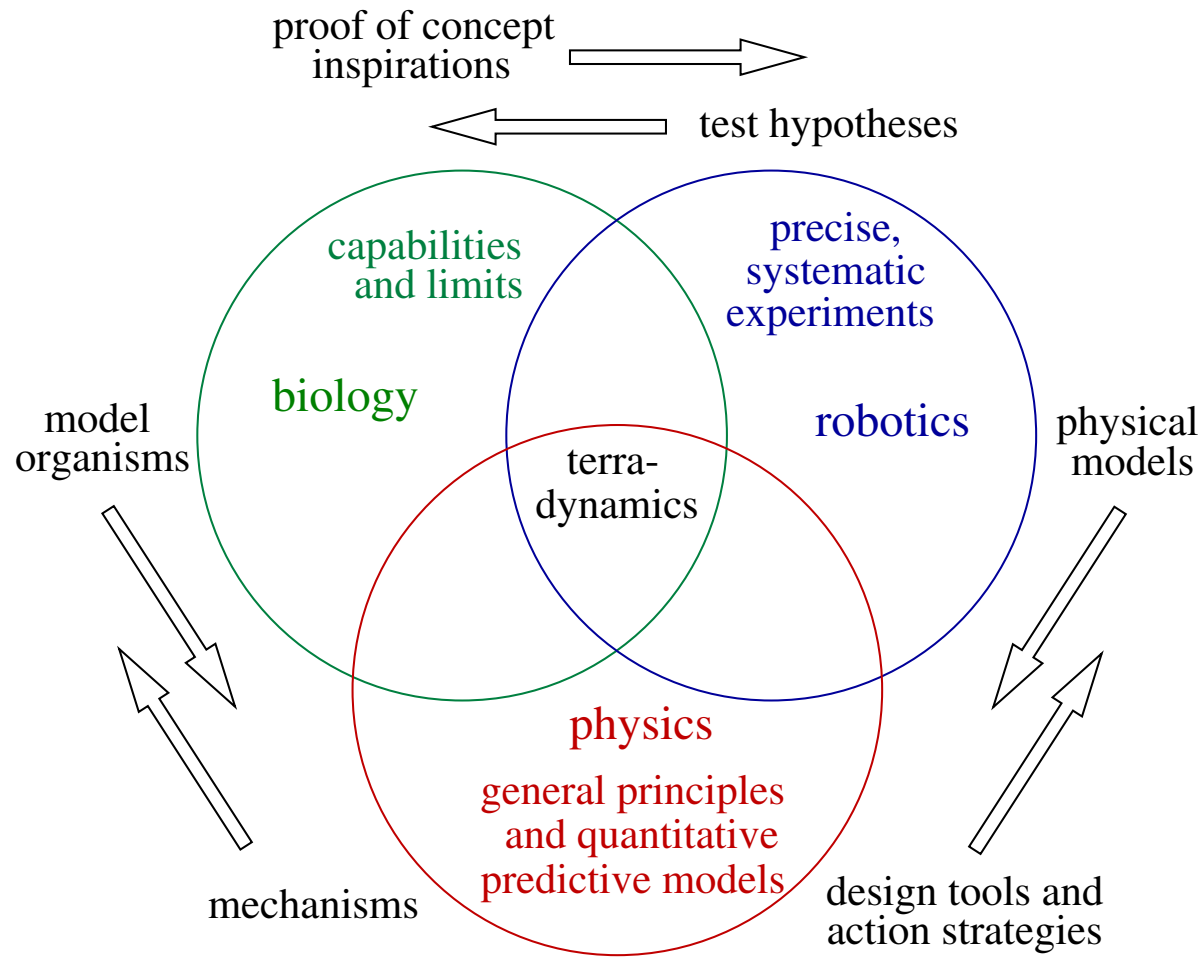

**Figure 2.** Integrative approach. Observations of model organisms inspire robot design and action. Simplified robots serve as physical models for testing biological hypotheses or generating new ones [28,29,41] and allow control and variation of parameters to discover general principles. Physical principles and predictive models from this empirical approach provide mechanistic explanations for animal locomotion and design tools and action strategies for robots. (Online version in colour.)

Although studying locomotor transitions to overcome these challenges separately is an amenable first step (figure 1*f*), in the real world, animals and robots must continually transition across locomotor modes to traverse diverse obstacles over large spatio-temporal scales (figure 1*e*). To study continual transitions, we developed a terrain treadmill (figure 3*e*) to study locomotion through large obstacles over a long time and a large distance [48], while allowing finer features such as antenna and leg motion to be observed at a high spatial resolution [49]. This research direction is still at an early stage.

## (b) Robotic physical models

We created simplified robotic physical models [28,29] of each model system (figure 3*f*–*j*). These robots offer several advantages as experimental platforms. First, they generate relevant locomotor behaviour using minimalistic design, actuation and sensing, facilitating analysis and modelling. In addition, they are more amenable than animals to controlled parameter variation and hypothesis testing. Moreover, running the robot in open loop allows isolating the effects of passive mechanics from that of sensory feedback. Finally, they cannot violate the laws of physics because robots are enacting, not modelling, the laws of physics [62].

We emphasize that our robots were designed and controlled to generate relevant locomotor transitions that we studied, not optimized for maximal performance. However, the physical principles revealed by these tools are generalizable and can predict how to increase performance [4,28,29,50–55,57–59] (§4d).

# 3. Modelling approaches

## (a) Potential energy landscape modelling

Understanding how locomotor transitions emerge from locomotor–terrain interaction probabilistically (§4a) calls for a statistical physics approach. A statistical physics treatment has advanced understanding of complex, stochastic, macroscopic phenomena in self-propelled living systems, such as animal foraging [63], traffic [64] and active matter [65,66]. Here, we created potential energy landscape models (figure 4*b*), directly inspired by free energy landscapes for modelling multi-pathway protein folding transitions [67–69]. The near-equilibrium, microscopic proteins statistically transition from higher to lower, thermodynamically more favourable states on the free energy landscape. Thermal fluctuation comparable to free energy barriers induces probabilistic barrier crossings. These physical principles operating on a rugged landscape leads to multi-pathway protein folding transitions. Although our locomotor–terrain interaction systems are macroscopic, self-propelled and far-from-equilibrium, their locomotor transitions display similar features, including stochasticity, multi-pathway transitions, kinetic energy fluctuation (from oscillatory self-propulsion) and favourability of some modes over others [4,51–54,56–59], but with the addition of intelligence.

Given these similarities, we hypothesized that locomotor transitions are barrier-crossing transitions between basins of potential energy landscapes of our systems. We tested this hypothesis in each model system (figure 4; electronic supplementary material, text S3–S7). To discover general principles of locomotor transitions, we systematically varied system parameters and studied how they affect locomotor transitions. For how to use potential energy landscape modelling, see electronic supplementary material, text S2.

A potential energy landscape approach to modelling locomotor–terrain interaction is plausible also considering the success of potential energy field methods in modelling robotic manipulation. Similar to our systems, robotic part alignment [70] and grasping [71] have continual collisions, multiple pathways to reach the goal [70] and favourability of some contact configurations over others [72]. Given these complexities, quasi-static potential energy fields well explained how system properties like geometry and friction affect part-manipulator interaction and informed strategies to achieve desired alignment or manipulation [70].

We emphasize that our potential energy landscapes directly result from physical interaction and are based on first principles, unlike artificially defined potential functions to explain walk-to-run transition [73] and other non-equilibrium biological phase transitions [74], metabolic energy landscapes inferred from oxygen consumption measurements to explain behavioural switching of locomotor modes [19] and artificial potential fields for robot obstacle avoidance [75].

For simplicity, our potential energy landscapes so far only considered the most relevant system degrees of freedom (body rotation and translation in obstacle traversal, body rotation and wing opening in self-righting). In addition, they do not yet model system dynamics, which is required for the quantitative prediction of locomotor transitions (§5a). Despite these limitations, they provided substantial insight into the general principles and strategies of obstacle traversal and strenuous ground self-righting (§4).

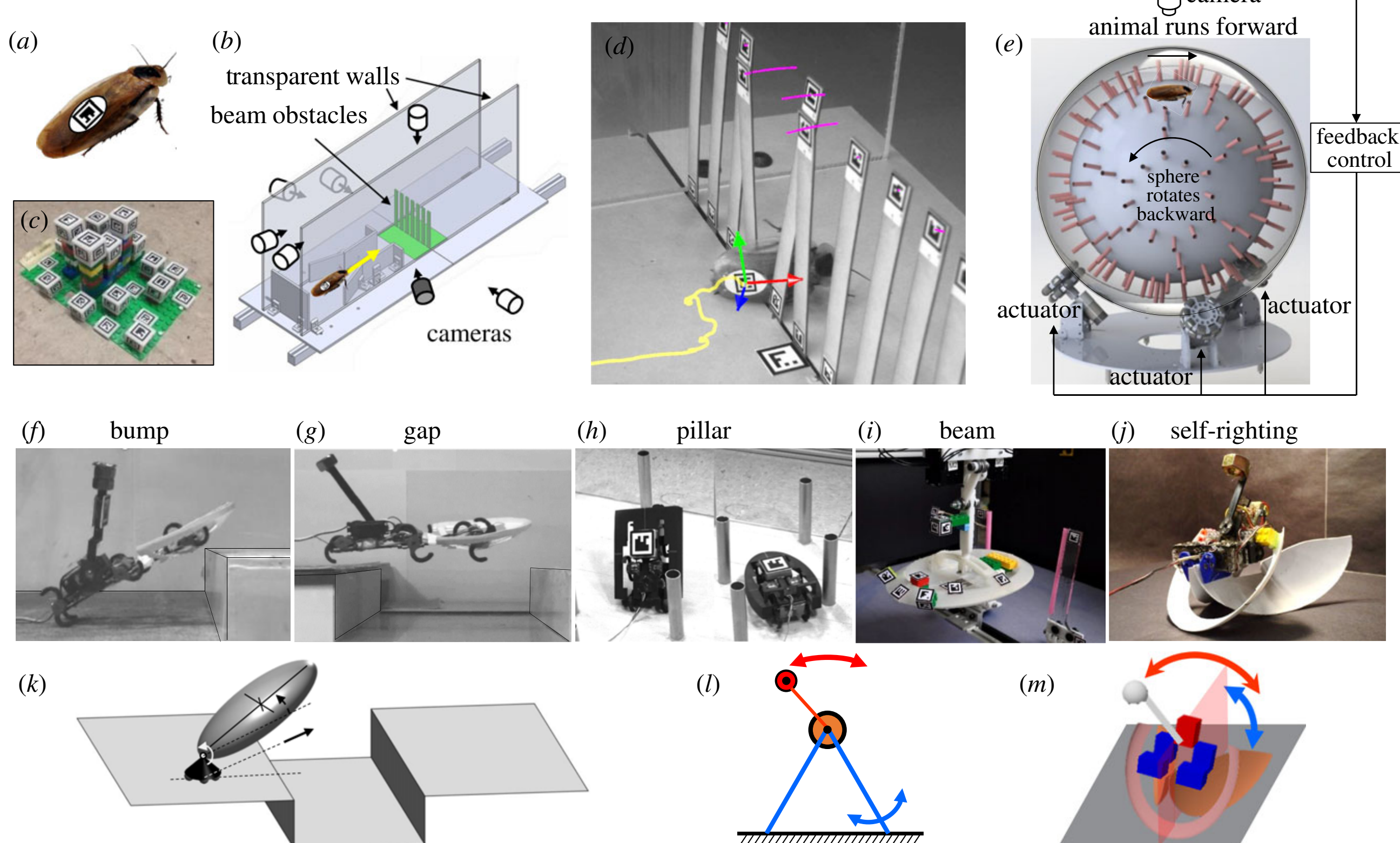

**Figure 3.** Experimental tools and dynamic models. (*a*) Model organism. (*b*) Terrain testbed with multi-camera imaging system. (*c*) Automated three-dimensional calibration object. (*d*) Snapshot of obstacle traversal showing automatically tracked trajectories of animal (yellow) and terrain (pink) markers. (*e*) Terrain treadmill with an untethered animal kept atop by rotating the sphere at the opposite velocity [48,49]. (*f–j*) Robotic physical models [4,50–59]. (*k,l*) Dynamical templates [53,58]. (*m*) Multi-body dynamics simulation [59]. (Online version in colour.)

### (b) Dynamic templates and simulations

Although our model systems follow Newton's laws, it is often challenging to solve equations of motion analytically due to the hybrid contact [22] and high-dimensional parameter space. As a first step to understand transition dynamics, we developed dynamical templates for two model systems, large gap traversal [53] (figure 3*k*) and strenuous ground self-righting [58] (figure 3*l*), for which equations of motion are solvable when two-dimensional dynamics is considered. Templates are the simplest dynamical models that capture the fundamental dynamics of a locomotor behaviour using minimal degrees of freedom [76]. For these two systems, our templates enabled quantitative prediction of contact and actuator forces [58], control strategies for traversal [53] or self-righting [58], and how they depend on system parameters [53,58].

In addition, for strenuous ground self-righting, we developed multi-body dynamics simulations of the robot validated against experiments [59] to study the effect of randomness in wing–leg coordination (figure 3*m*). These simulations enabled large-scale variation of relevant parameters identified from experiments and in-depth analysis at a precision difficult to achieve in animals and robots. Finally, simulation is faster than experiments [59].

## 4. Insights and general principles from simple model systems

Our studies revealed how locomotor transitions depend on system parameters (gap width, beam stiffness, body shape, etc.; electronic supplementary material, table S1). For each model system, these physical principles are generalizable over the relevant parameter space and helped improve robot performance. Although our model systems are level, our approach also applies to interactions on slopes.

Across model systems, a potential energy landscape approach helps understand how the animal's and robot's stereotyped, probabilistic locomotor transitions are constrained by physical interaction. Several general physical principles and new concepts emerge.

### (a) Locomotor modes are stereotypical and transitions are stochastic

For all model systems, the animal displayed stereotyped locomotor modes with qualitatively similar body postural changes [4,50–54,56,57]. Not all modes lead to successful obstacle traversal or self-righting. Transitions between modes occur stochastically, with large trial-to-trial variation [4,50,51,53,54,56,57]. The probability of using or transitioning to a mode strongly depends on locomotor and terrain parameters that affect physical interaction [4,50,52–54,56,57] (§4f). The robot's locomotor modes are also stereotyped and transitions stochastic [4,50–55,57].

### (b) Locomotor transitions are destabilizing barrier-crossing transitions on a potential energy landscape

For all model systems, the system state in each mode is strongly attracted to a local minimum basin of the potential energy landscape over the relevant body state space [50–52,54,57] (figure 4; electronic supplementary material, figures S2–S6 and movie S1). This is because self-propulsion induces continual body–terrain collisions during obstacle interaction and self-righting, which breaks continuous frictional contact and makes the system statically unstable. This leads the system to drift down the basin until a sufficient perturbation induces an escape from the basin. However, the system does not stay at the minimum due to self-propulsion. Due to this strong attraction to landscape basins, the transition from one locomotor mode to another requires the system to destabilize itself to escape from one basin to fall into another.

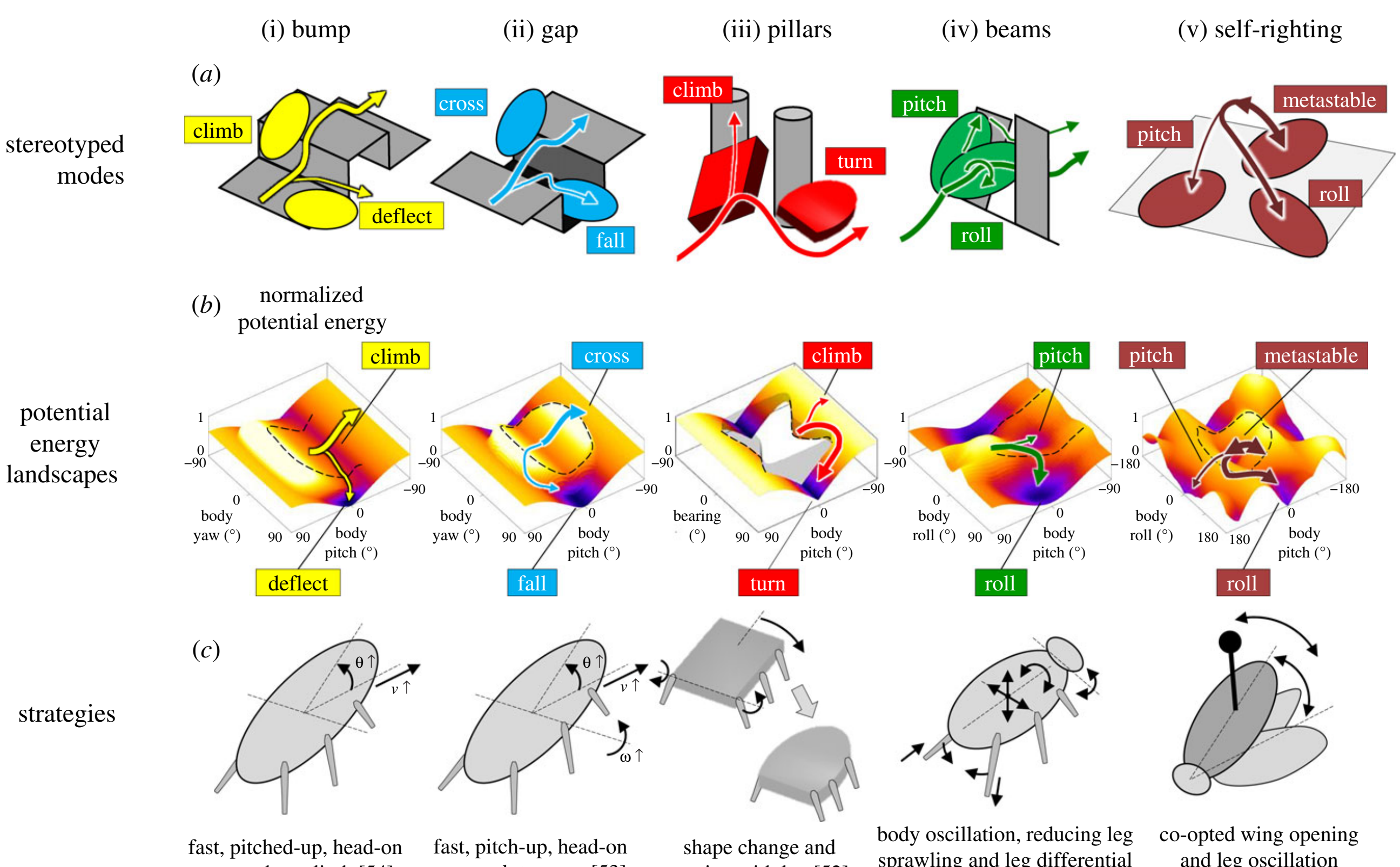

**Figure 4.** Modulation of locomotor transitions on potential energy landscapes via a suite of strategies. (*a*) Stereotyped locomotor modes of model systems. (*b*) Potential energy landscapes. The system is attracted to a distinct basin in each mode. A potential energy barrier must be crossed to make locomotor transitions. Black dashed curves show potential energy barriers. Arrows in (*a*,*b*) show representative system state trajectories; thicker arrows show more desirable modes. (*c*) Strategies that can increase the probabilities of desired modes and facilitate transitions to overcome locomotor challenges. (i–v) Model systems. See electronic supplementary material, table S1 and text S3–S7 for detail. We renamed some modes/basins in this review from in the original papers to better distinguish them across model systems. (Online version in colour.)

## (c) There exists a potential energy landscape-dominated regime of locomotion

These observations across diverse model systems mean that there is a potential energy landscape-dominated regime of locomotion. In this regime, along with certain directions, there exist large potential energy barriers that are comparable to or exceed kinetic energy and/or mechanical work generated by each propulsive cycle or motion. This may happen when propulsive forces are either limited by physiological, morphological and environmental (e.g. low friction) constraints or are not well directed towards directions along which large barriers exist for the desired transition. These situations are frequent in large obstacle traversal and strenuous ground self-righting. In this regime, not only do potential energy landscapes provide a useful statistical physics approach for understanding locomotor transitions, but it also allows comparison across systems (different species [56], robots [4,52], terrain [50,52–54] and modes [4,50,52,56,57]) to discover general principles. Outside of this regime, potential energy landscapes are not useful or necessary. Such examples include ballistic jumping over small obstacles with kinetic energy far exceeding potential energy barriers, moving on slopes with potential energy increasing or decreasing monotonically, and traversing obstacles much smaller or larger than body size.

## (d) Feed-forward self-propulsion can induce locomotor transitions

Using robotic physical models, we discovered several principles of locomotor transitions with feed-forward self-propulsion. First, locomotor kinetic energy fluctuation from self-propulsion helps the system stochastically cross potential energy barriers to make transitions [50,57]. In addition, escape from a basin is more likely in directions on the landscape along which the barriers are lower [50,57]. Finally, during a transition, the system tends to transition to more favourable modes attracted to lower basins [50,52,57]. The animal's locomotor transitions also largely followed these principles during rapid, bandwidth-limited escape or emergency self-righting response [50–54,56,57].

## (e) Feedback-controlled active adjustments can assist locomotor transitions

Not surprisingly, the animal can make active adjustments to facilitate or enable desired transitions when feed-forward self-propulsion is insufficient. For example, even when body kinetic energy fluctuation becomes comparable to, but is still lower than, the potential energy barrier, the animal transitions to a more favourable mode to traverse beam obstacles [50], by actively adjusting body and appendages [51]. Understanding this intelligence aspect of locomotor transitions is clearly the next step. We have begun studying the principles of feedback-controlled locomotor transitions by creating robotic physical models with force sensing [51].

### (f) A suite of strategies can modulate locomotor transitions and increase performance

Because locomotor transitions are barrier-crossing transitions, they can be enhanced or suppressed by steering the system state on the landscape, changing landscape barriers, or even modifying landscape topology (the number of basins). This insight allowed us to discover a suite of strategies (figure 4*c*) to make desired transitions more probable for each model system (figure 4*a*), elaborated below.

In bump traversal, approaching with a head-on (body sagittal plane perpendicular to bump), pitched-up body posture directs the system to overcome a barrier to reach a desired climb basin/mode and avoid being attracted towards a deflect basin/mode (figure 4(i)) [54]. Similarly, in gap traversal, approaching with a large forward velocity and upward pitching velocity and a head-on, pitched-up body posture increases kinetic energy that directs the system to reach a desired cross basin/mode and avoid being attracted into a fall basin/mode (figure 4(ii)) [53].

In pillar traversal, a cuboidal body induces a climb basin/mode where the body is attracted to and pitches up against the pillar, whereas an elliptical body eliminates it and induces a desirable turn basin/mode where the body is repelled away (figure 4(iii)) [52]. Alternatively, active turning by legs helps a cuboidal body steer away from the climb basin/mode and cross the barrier to transition to the turn basin/mode [52]. In beam traversal, when beams are stiff, it is challenging to push across in a pitched-up mode attracted to a pitch basin, and it is desirable to transition to a roll mode/basin to roll into the beam gap to traverse (figure 4(iv)). Body kinetic energy fluctuation from self-propulsion helps cross the barrier to make this transition [50]. This transition is further facilitated by reducing sprawling and differential use of hind legs, which presumably destabilize and steer the system towards the roll basin [51].

In strenuous ground self-righting (figure 4(v)), although wing opening initiates a somersault and steers the system towards an upright pitch basin/mode, it is insufficient to overcome the large barrier. As a result, the system is frequently trapped in a metastable basin/mode due to a triangular base of support, leading to repeated failed attempts. However, wing opening reduces the barrier to transition from the metastable to a roll basin/mode, allowing small kinetic energy fluctuation from leg oscillation to induce barrier crossing, resulting in self-righting by rolling [57]. This transition is also facilitated by proper wing–leg coordination that better steers the system towards the lowered barrier to roll [58]. Randomness in wing–leg coordination helps find proper coordination [59].

We emphasize that the desirable modes and strategies in the obstacle interactions above aim at successful traversal. In different tasks, other modes may be desirable. For example, the fall mode in gap interaction (figure 4*a*,*b*(ii)) is desirable for going into ground crevices, and the climb mode for pillar interaction (figure 4*a*,*b*(iii)) is desirable for initiating climbing up obstacles. Strategies can be discovered for these modes accordingly using the same approach.

Using our feed-forward-controlled robotic physical models [50,52,57–59] or with a human in the loop to switch on the strategies [52–54,58], we have demonstrated that these strategies increased robot performance substantially or even enabled new capabilities in each model system (electronic supplementary material, table S1). Efforts remain to study how robots should sense locomotor–terrain interaction and use feedback control to make transitions intelligently.

### (g) Stereotyped locomotor modes result from physical interaction constraint

Although the self-propelled system can in principle move in arbitrary ways, the observed locomotor modes are highly stereotyped due to strong constraints from physical interaction (§4a). This stereotypy is because the potential energy landscape is highly rugged, with distinct basins separated by barriers, and the system is strongly attracted to landscape basins in the potential energy landscape-dominated regime. Because our potential energy landscapes are directly derived from first principles (as opposed to fitting a model to behavioural data [77,78]), this insight provided evidence that behavioural stereotypy of animals emerges from the physical interaction of their neural and mechanical systems with the environment [12,13]. In addition, our systematic studies revealed that variation in movement can lead to stochastic locomotor transitions and is advantageous when locomotor behaviour is separated into distinct modes, each of which may be desirable for different scenarios.

We speculate that this physical constraint plays a role in the evolution of animal morphology and behaviour. This is plausible because morphological [79–81] and behavioural [82] adaptations that facilitate obstacle traversal and self-righting are common when microhabitat properties physically constrain movement. Our potential energy landscape approach is also useful for quantifying how physical interaction constrains robot design, control and planning for locomotor transitions in the large locomotor and terrain parameter space.

### (h) Physical principles of locomotor–terrain interaction are general

In the potential energy landscape-dominated regime, physical principles and strategies that we discovered (figure 4*c*; electronic supplementary material, table S1) are applicable to a broad range of the parameter space of model systems. For example, obstacle attraction or repulsion is an inherent property of the locomotor shape and insensitive to pillar size and geometry [52]. Strategies that favour bump or gap traversal are applicable to a large range of bump heights [54] or gaps widths [53]. Physical principles of beam interaction explained how pitch-to-roll transition probability changes as beam stiffness varies over a large range [50].

## 5. Towards multi-pathway locomotor transitions

Considering the general physical principles of locomotor transitions from diverse simple model systems, we hypothesize that multi-pathway locomotor transitions in heterogeneous complex three-dimensional terrains can be understood by composing larger-scale, higher-dimensional potential energy landscapes (figure 5) from simple landscapes of abstracted challenges (e.g. figure 1*d–f*). Our terrain treadmill experiments (figure 3*e*) are beginning to shed light on this [49]. Progress towards such an understanding will lead to advancement for several fields.

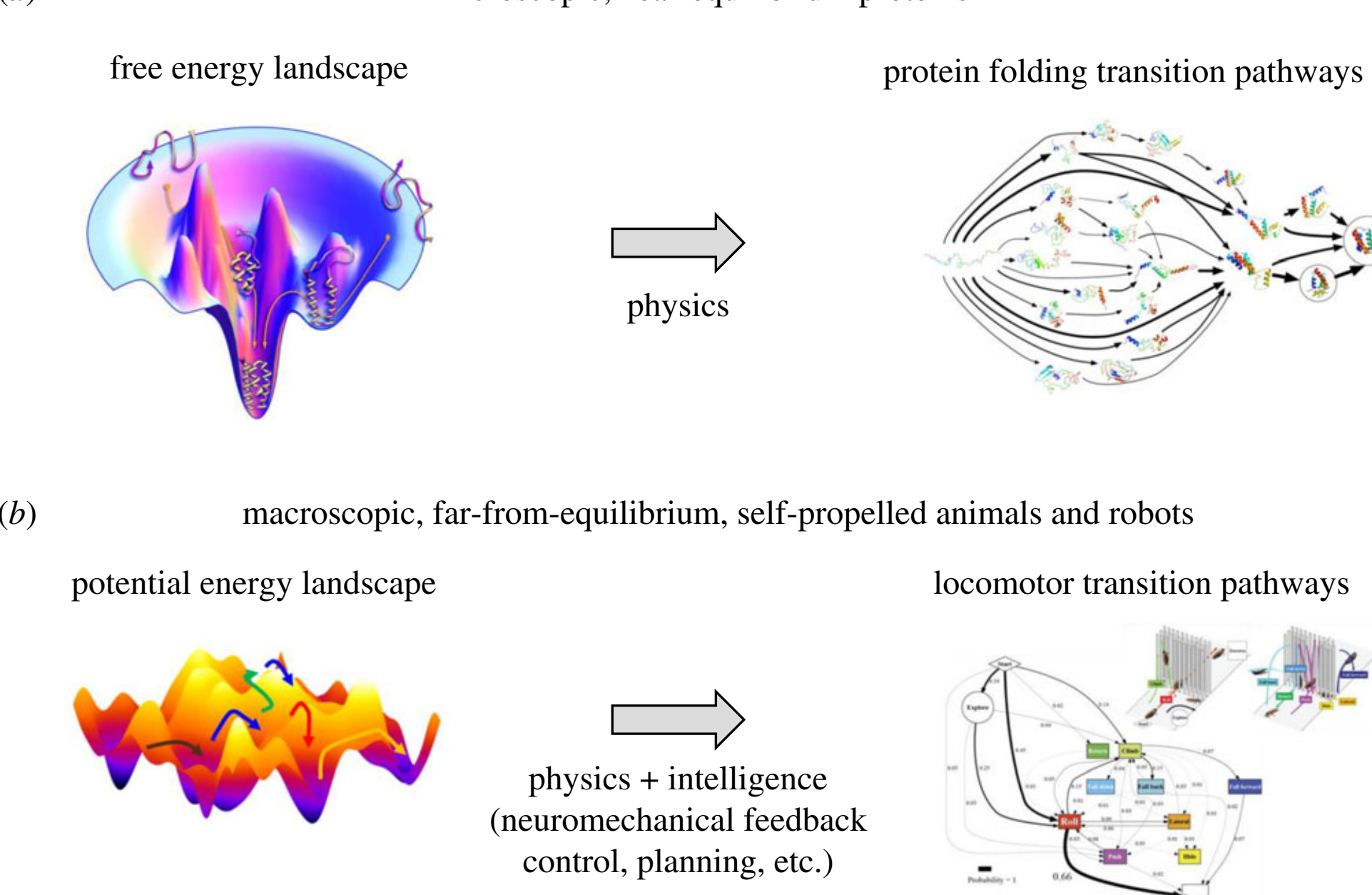

**Figure 5.** Comparison of two energy landscape approaches. (*a*) Rugged free energy landscapes help understand how proteins fold to their native states by stochastically transitioning from higher to lower free energy states via multiple pathways [67–69]. (*b*) We envision energy landscape modeling as a beginning of a statistical physics approach, but with the addition of intelligence, for understanding how the neuromechanical control system mediates physical interaction to generate multi-pathway locomotor transitions in complex 3-D terrain. Note that our locomotor-terrain interaction system differs from protein folding in that animals and robots are macroscopic, self-propelled, far-from-equilibrium and can have intelligence. Image credits: (*a*) Left: from [83]. Reprinted with permission from AAAS. Right: adapted with permission from [84]. Copyright © (2012) American Chemical Society. (*b*) Right: copyright © IOP Publishing. Reproduced with permission from [4]. All rights reserved.

## (a) Envisioned advancement for physics

The empirically discovered physical principles of locomotor transitions using feed-forward self-propulsion (§4d) are surprisingly similar to those of microscopic multi-pathway protein folding transitions (see detail in [50]), where predictive free energy landscape theories have been successful [67–69]. This was unexpected, given the differences in scale and nature of the interaction (macroscopic contact forces in locomotion versus ionic and dipole interactions, hydrogen bonds, van der Waals forces, hydrophobic interactions in protein folding) [68].

We envision the creation of analogous potential energy landscape theories, but with the addition of intelligence (e.g. §4e,f), to understand and predict how the animal's nervous system or robot's sensing, control and planning systems mediate physical interaction to generate multi-pathway locomotor transitions (such as observed in [4]). The next step towards this is to model conservative forces using potential energy landscape gradients, add stochastic, non-conservative propulsive and dissipative forces that perturb the system to 'diffuse' across landscape barriers (analogous to [85], but with closed-loop control of the landscape over locomotor degrees of freedom), and simulate multi-pathway locomotor transitions. Systematic studies to understand the principles of force sensing [51] will inform how to steer the system and modify the landscape to modulate transitions intelligently using sensory feedback control. Such new theories will help expand the physics of living systems to the organismal level and expand statistical physics to macroscopic, far-from-equilibrium, self-propelled (active) systems [65,66].

## (b) Envisioned advancement for dynamical systems theory

Our potential energy landscape approach provided a new conceptual way of thinking about locomotor modes beyond near-steady-state, limit cycle-like behaviour (e.g. walk, run and climb [5–7]) (electronic supplementary material, figure S8*a*). Locomotion in irregular terrain with repeated perturbations requires an animal to continually modify its behaviour, which cannot be described by limit cycles [61]. Our work demonstrated that, in the potential energy landscape-dominated regime, the system must destabilize from an attractive landscape basin to transition from one mode to another, and locomotor modes can be metastable [86], far-from-steady manoeuvers (e.g. electronic supplementary material, figure S8*c*). We foresee the creation of new dynamical systems theories of terrestrial locomotion [22] that are composed of multi-pathway transitions across modes attracted to both landscape basins attractors and limit cycles [87] (electronic supplementary material, figure S8*d*).

In addition, such new dynamical systems theories modelling physical interaction may be combined with those that model related processes and factors such as proprioception [88], external sensory cues (e.g. predators, prey, resources) [14,89], internal needs (e.g. hunger, mating) [90] and safety–risk tradeoffs [91]. This integration will elucidate how these factors interplay with physical interaction to modulate animals' locomotor transition behaviour in complex environments.

### (c) Envisioned advancement for biology

Our potential energy landscape approach provides a means towards the first principle, physical understanding of the organization of locomotor behaviour, filling a critical knowledge gap. The field of movement ecology [14] makes field observations of trajectories of animals—often as a point mass (e.g. [92])—moving and making behavioural transitions in natural environments, because physical interactions are difficult to measure at such large scales. Recent progress in quantitative ethology advanced understanding of the organization of behaviours, often by quantifying kinematics in homogeneous, near-featureless laboratory environments (see [12,13] for reviews). Our work highlights the importance and feasibility of, and opens new avenues for, studying how stereotypy and organization of behaviour are constrained by an animal's direct physical interaction with realistic environments. Analysing the disconnectivity [69] of basins of future composed landscapes for multi-pathway transitions will reveal the hierarchy ('treeness' [93]) of locomotor modes.

In addition, there are opportunities to explore how physical interaction during locomotion impacts large-scale processes like predator–prey pursuit and migration where locomotor performance is crucial [94]. If future potential energy landscape theories can predict how locomotor performance depends on relevant system parameters (§4d–g), they will provide a proxy for fitness landscapes [95]. Such proxy fitness landscapes will reveal how locomotor fitness exerts selective pressure on morphology and behaviour that affect locomotor transitions via physical interaction.

### (d) Envisioned advancement for robotics

Future predictive potential energy landscape theories will predict strategies for robots to use physical interaction to generate landscape basin attractors funnelled into one another [96] to compose locomotor transitions to perform high-level, goal-directed tasks in the real world. Using information of the geometry and physical properties of complex three-dimensional terrain from sensors, a robot can abstract its locomotor task into separate locomotor challenges (figure 1*e*) and calculate their potential energy landscapes. Then, the robot can use the landscape theories to identify possible transitions (figure 1*d*) and predict how transition probabilities differ between strategies (figure 4*a*,*b*). Finally, within its own constraints (e.g. energy available and actuator force limits), the robot can plan its strategies to make transitions that increase or even optimize its probability to reach the goal (figure 1*d*). When the terrain is sensed only up to a finite horizon with uncertainty, the robot can react to newly sensed challenges or recently failed attempts and update the pre-planned locomotor transition sequence and strategies (analogous to reactive obstacle avoidance using geometry [97]).

Recent learning approaches have managed to generate slow locomotion where terrain perturbations are sufficiently small for the learned controller to reject and stabilize the robot around an upright body posture [98,99]. Although learning approaches can in principle train the robot for any task in simulation by brute force, even in such modest terrain, the real system's physics must still be modelled properly (e.g. how motor dynamics affects leg dynamics) to narrow the simulation-to-reality gap [98,99]. However, as our work reveals, a robot should use physical interaction to destabilize itself to make locomotor transitions to traverse large obstacles. In addition, locomotor transitions are diverse and stochastic, and they depend sensitively on locomotor and terrain parameters and vary substantially with strategies. Considering these, learning approaches alone will be fragile for generating robot locomotor transitions in complex three-dimensional terrain. Our physics approach will be crucial for applying learning approaches here—it not only enables robots with basic transition capabilities (§4f; electronic supplementary material, table S1) to serve as real platforms for learning, but also offers principles of how strategies affect transitions across the large locomotor and terrain parameter space (§4g) to guide learning.

In the longer term, we envision that first principle models of locomotor–terrain physical interaction will be pervasive. Analogous to self-driving cars that scan streets, robots will create environmental physics maps and action databases for locomotor transitions and add them to geometric maps in the cloud for shared use [100]. They will help robots better use physical interaction to traverse currently unreachable complex three-dimensional terrain and expand our reaches in natural, artificial and extraterrestrial terrain.

Data accessibility. Data and code are available at https://github.com/TerradynamicsLab/potential_energy_landscape. An overview video is available at https://doi.org/10.6084/m9.figshare.14207927.

Authors' contributions. R.O. created visualizations and wrote the paper, Q.X. and Y.W. provided feedback, and C.L. revised the paper.

Competing interests. The authors declare no competing interests.

Funding. This work was funded by an Army Research Office Young Investigator Program (grant no. W911NF-17-1-0346), a Burroughs Wellcome Fund Career Award at the Scientific Interface, an Arnold and Mabel Beckman Foundation Beckman Young Investigator award and The Johns Hopkins University Whiting School of Engineering start-up funds to C.L.

Acknowledgements. We thank previous members of the Terradynamics lab, especially Sean Gart and Yuanfeng Han, for contribution to the studies reviewed here, Dan Goldman, Simon Sponberg, Bob Full, Dan Koditschek, Shai Revzen and Noah Cowan for discussion, and three anonymous reviewers for suggestions.

## References

1. Dickinson MH, Farley CT, Full RJ, Koehl MAR, Kram R, Lehman S. 2000 How animals move: an integrative view. *Science* **288**, 100–106. (doi:10.1126/science.288.5463.100)
2. Alexander RM. 2006 *Principles of animal locomotion*. Princeton, NJ: Princeton University Press.
3. Lock RJ, Burgess SC, Vaidyanathan R. 2013 Multi-modal locomotion: from animal to application. *Bioinspir. Biomim.* **9**, 011001. (doi:10.1088/1748-3182/9/1/011001)
4. Li C, Pullin AO, Haldane DW, Lam HK, Fearing RS, Full RJ. 2015 Terradynamically streamlined shapes in animals and robots enhance traversability through densely cluttered terrain. *Bioinspir. Biomim.* **10**, 46003. (doi:10.1088/1748-3190/10/4/046003)
5. Blickhan R, Full RJ. 1993 Similarity in multilegged locomotion: bouncing like a monopode. *J. Comp. Physiol. A* **173**, 509–517. (doi:10.1007/BF00197760)

6. Kuo AD. 2007 The six determinants of gait and the inverted pendulum analogy: a dynamic walking perspective. *Hum. Mov. Sci.* **26**, 617–656. (doi:10.1016/j.humov.2007.04.003)
7. Goldman DI, Chen TS, Dudek DM, Full RJ. 2006 Dynamics of rapid vertical climbing in cockroaches reveals a template. *J. Exp. Biol.* **209**, 2990–3000. (doi:10.1242/jeb.02322)
8. Hu DL, Nirody J, Scott T, Shelley MJ. 2009 The mechanics of slithering locomotion. *Proc. Natl Acad. Sci. USA* **106**, 10 081–10 085. (doi:10.1073/pnas.0812533106)
9. Biewener AA, Daley MA. 2007 Unsteady locomotion: integrating muscle function with whole body dynamics and neuromuscular control. *J. Exp. Biol.* **210**, 2949–2960. (doi:10.1242/jeb.005801)
10. Revzen S, Burden SA, Moore TY, Mongeau JM, Full RJ. 2013 Instantaneous kinematic phase reflects neuromechanical response to lateral perturbations of running cockroaches. *Biol. Cybern.* **107**, 179–200. (doi:10.1007/s00422-012-0545-z)
11. Birn-Jeffery A V, Hubicki CM, Blum Y, Renjewski D, Hurst JW, Daley MA. 2014 Don't break a leg: running birds from quail to ostrich prioritise leg safety and economy on uneven terrain. *J. Exp. Biol.* **217**, 3786–3796. (doi:10.1242/jeb.102640)
12. Berman GJ. 2018 Measuring behavior across scales. *BMC Biol.* **16**, 23. (doi:10.1186/s12915-018-0494-7)
13. Brown AEX, de Bivort B. 2018 Ethology as a physical science. *Nat. Phys.* **14**, 653–657. (doi:10.1038/s41567-018-0093-0)
14. Nathan R, Getz WM, Revilla E, Holyoak M, Kadmon R, Saltz D, Smouse PE. 2008 A movement ecology paradigm for unifying organismal movement research. *Proc. Natl Acad. Sci. USA* **105**, 19 052–19 059. (doi:10.1073/pnas.0800375105)
15. Ijspeert AJ. 2008 Central pattern generators for locomotion control in animals and robots: a review. *Neural Netw.* **21**, 642–653. (doi:10.1016/j.neunet.2008.03.014)
16. Blaesing B, Cruse H. 2004 Stick insect locomotion in a complex environment: climbing over large gaps. *J. Exp. Biol.* **207**, 1273–1286. (doi:10.1242/jeb.00888)
17. Kohlsdorf T, Biewener AA. 2006 Negotiating obstacles: running kinematics of the lizard *Sceloporus malachiticus*. *J. Zool.* **270**, 359–371. (doi:10.1111/j.1469-7998.2006.00150.x)
18. Ritzmann RE *et al.* 2012 Deciding which way to go: how do insects alter movements to negotiate barriers? *Front. Neurosci.* **6**, 1–10. (doi:10.3389/fnins.2012.00097)
19. Shepard ELC, Wilson RP, Rees WG, Grundy E, Lambertucci SA, Vosper SB. 2013 Energy landscapes shape animal movement ecology. *Am. Nat.* **182**, 298–312. (doi:10.1086/671257)
20. Bramble DM, Lieberman DE. 2004 Endurance running and the evolution of *Homo*. *Nature* **432**, 345–352. (doi:10.1038/nature03052)
21. Li C, Zhang T, Goldman DI. 2013 A terradynamics of legged locomotion on granular media. *Science* **339**, 1408–1412. (doi:10.1126/science.1229163)
22. Holmes P, Full RJ, Koditschek D, Guckenheimer J. 2006 The dynamics of legged locomotion: models, analyses, and challenges. *SIAM Rev.* **48**, 207–304. (doi:10.1137/S0036144504445133)
23. Kumar GA, Patil AK, Patil R, Park SS, Chai YH. 2017 A LiDAR and IMU integrated indoor navigation system for UAVs and its application in real-time pipeline classification. *Sensors* **17**(6), 1268. (doi:10.3390/s17061268)
24. Thrun S. 2010 Toward robotic cars. *Commun. ACM* **53**, 99. (doi:10.1145/1721654.1721679)
25. Pacejka HB. 2005 *Tyre and vehicle dynamics*. Amsterdam, The Netherlands: Elsevier.
26. Wong JY. 2009 *Terramechanics and off-road vehicle engineering: terrain behaviour, off-road vehicle performance and design*. Amsterdam, The Netherlands: Elsevier.
27. Vogel S. 1996 *Life in moving fluids: the physical biology of flow*. Princeton, NJ: Princeton University Press.
28. Aguilar J *et al.* 2016 A review on locomotion robophysics: the study of movement at the intersection of robotics, soft matter and dynamical systems. *Rep. Prog. Phys.* **79**, 110001. (doi:10.1088/0034-4885/79/11/110001)
29. Ijspeert AJ. 2014 Biorobotics: using robots to emulate and investigate agile locomotion. *Science* **346**, 196–203. (doi:10.1126/science.1254486)
30. Goldman DI. 2014 Colloquium: biophysical principles of undulatory self-propulsion in granular media. *Rev. Mod. Phys.* **86**, 943–958. (doi:10.1103/RevModPhys.86.943)
31. Maladen RD, Umbanhowar PB, Ding Y, Masse A, Goldman DI. 2011 Granular lift forces predict vertical motion of a sand-swimming robot. *Proc. IEEE Int. Conf. Robot. Autom.* 1398–1403. (doi:10.1109/ICRA.2011.5980301)
32. Li C, Hsieh T, Goldman D. 2012 Multi-functional foot use during running in the zebra-tailed lizard (*Callisaurus draconoides*). *J. Exp. Biol.* **215**, 3293–3308. (doi:10.1242/jeb.061937)
33. Sharpe SS, Koehler SA, Kuckuk RM, Serrano M, Vela PA, Mendelson J, Goldman DI. 2015 Locomotor benefits of being a slender and slick sand swimmer. *J. Exp. Biol.* **218**, 440–450. (doi:10.1242/jeb.108357)
34. Sharpe SS, Ding Y, Goldman DI. 2013 Environmental interaction influences muscle activation strategy during sand-swimming in the sandfish lizard *Scincus scincus*. *J. Exp. Biol.* **216**, 260–274. (doi:10.1242/jeb.070482)
35. Ding Y, Sharpe SS, Wiesenfeld K, Goldman DI. 2013 Emergence of the advancing neuromechanical phase in a resistive force dominated medium. *Proc. Natl Acad. Sci. USA* **110**, 10 123–10 128. (doi:10.1073/pnas.1302844110)
36. McInroe B, Astley HC, Gong C, Kawano SM, Schiebel PE, Rieser JM, Choset H, Blob RW, Goldman DI. 2016 Tail use improves performance on soft substrates in models of early vertebrate land locomotors. *Science* **353**, 154–158. (doi:10.1126/science.aaf0984)
37. Li C, Umbanhowar PB, Komsuoglu H, Koditschek DE, Goldman DI. 2009 Sensitive dependence of the motion of a legged robot on granular media. *Proc. Natl Acad. Sci. USA* **106**, 3029–3034. (doi:10.1073/pnas.0809095106)
38. Shrivastava S, Karsai A, Aydin YO, Pettinger R, Bluethmann W, Ambrose RO, Goldman DI. 2020 Material remodeling and unconventional gaits facilitate locomotion of a robophysical rover over granular terrain. *Sci. Robot.* **5**, eaba3499. (doi:10.1126/scirobotics.aba3499)
39. Marvi H *et al.* 2014 Sidewinding with minimal slip: snake and robot ascent of sandy slopes. *Science* **346**, 224–229. (doi:10.1126/science.1255718)
40. Li C, Umbanhowar PB, Komsuoglu H, Goldman DI. 2010 The effect of limb kinematics on the speed of a legged robot on granular media. *Exp. Mech.* **50**, 1383–1393. (doi:10.1007/S11340-010-9347-1)
41. Gravish N, Lauder GV. 2018 Robotics-inspired biology. *J. Exp. Biol.* **221**, jeb138438. (doi:10.1242/jeb.138438)
42. Gart SW, Mitchel TW, Li C. 2019 Snakes partition their body to traverse large steps stably. *J. Exp. Biol.* **222**, jeb185991. (doi:10.1242/jeb.185991)
43. Fu Q, Li C. 2020 Robotic modeling of snake traversing large, smooth obstacles reveals stability benefits of body compliance. *R. Soc. Open Sci.* **7**, 191192. (doi:10.1098/rsos.191192)
44. Fu Q, Mitchel TW, Kim JS, Chirikjian GS, Li C. 2021 Continuous body 3-D reconstruction of limbless animals. *J. Exp. Biol.* **224**, jeb220731. (doi:10.1242/jeb.220731)
45. Fu Q, Astley H, Li C. 2021 Snakes traversing complex 3-D terrain. *Integr. Comp. Biol.* **61**, e279.
46. Ramesh D, Fu Q, Wang K, Othayoth R, Li C. 2021 A sensorized robophysical model to study snake locomotion in complex 3-D terrain. *Integr. Comp. Biol.* **61**, e729–e730. (doi:10.1093/icb/icab001)
47. Fu Q, Gart SW, Mitchel TW, Kim JS, Chirikjian GS, Li C. 2020 Lateral oscillation and body compliance help snakes and snake robots stably traverse large, smooth obstacles. *Integr. Comp. Biol.* **60**, 171–179. (doi:10.1093/icb/icaa013)
48. Strebel B, Han Y, Li C. 2018 A novel terrain treadmill to study animal locomotion in complex 3-D terrains. *Integr. Comp. Biol.* **58**, e224. (doi:10.1093/icb/icy001)
49. Othayoth R, Francois E, Li C. 2021 Large spatiotemporal scale measurement of cockroach traversal of large obstacles. *Integr. Comp. Biol.* **61**, e673–e674.
50. Othayoth R, Thoms G, Li C. 2020 An energy landscape approach to locomotor transitions in complex 3D terrain. *Proc. Natl Acad. Sci. USA* **117**, 14 987–14 995. (doi:10.1073/pnas.1918297117)
51. Wang Y, Othayoth R, Li C. 2021 Uncovering the role of head flexion during beam obstacle traversal of cockroaches. *Integr. Comp. Biol.* **61**, e959.
52. Han Y, Othayoth R, Wang Y, Hsu C-C, de la Tijera Obert R, Francois E, Li C. In press. Shape-induced obstacle attraction and repulsion during dynamic locomotion. *Int. J. Rob. Res.* (doi:10.1177/0278364921989372)

53. Gart SW, Yan C, Othayoth R, Ren Z, Li C. 2018 Dynamic traversal of large gaps by insects and legged robots reveals a template. *Bioinspir. Biomim.* **13**, 026006. (doi:10.1088/1748-3190/aaa2cd)
54. Gart SW, Li C. 2018 Body-terrain interaction affects large bump traversal of insects and legged robots. *Bioinspir. Biomim.* **13**, 026005. (doi:10.1088/1748-3190/aaa2d0)
55. Li C, Kessens CC, Fearing RS, Full RJ. 2017 Mechanical principles of dynamic terrestrial self-righting using wings. *Adv. Robot.* **31**, 881–900. (doi:10.1080/01691864.2017.1372213)
56. Li C, Wöhrl T, Lam HK, Full RJ. 2019 Cockroaches use diverse strategies to self-right on the ground. *J. Exp. Biol.* **222**, jeb186080. (doi:10.1242/jeb.186080)
57. Othayoth R, Xuan Q, Li C. 2021 Co-opting propelling and perturbing appendages facilitates strenuous ground self-righting. *bioRxiv*. (doi:10.1101/2021.04.06.438657)
58. Xuan Q, Li C. 2020 Coordinated appendages accumulate more energy to self-right on the ground. *IEEE Robot. Autom. Lett.* **5**, 6137–6144. (doi:10.1109/LRA.2020.3011389)
59. Xuan Q, Li C. 2020 Randomness in appendage coordination facilitates strenuous ground self-righting. *Bioinspir. Biomim.* **21**, 1–9. (doi:10.1088/1748-3190/abac47)
60. Cayley G. 1876 On aërial navigation. *Annu. Rep. Aeronaut. Soc. Gt. Britain* **11**, 60–94. (doi:10.1017/S2397930500000552)
61. Sponberg S, Full RJ. 2008 Neuromechanical response of musculo-skeletal structures in cockroaches during rapid running on rough terrain. *J. Exp. Biol.* **211**, 433–446. (doi:10.1242/jeb.012385)
62. Long J. 2012 *Darwin's devices: what evolving robots can teach us about the history of life and the future of technology*. New York, NY: Basic Books.
63. Viswanathan GM, Da Luz MGE, Raposo EP, Stanley HE. 2011 *The physics of foraging*. Cambridge, UK: Cambridge University Press.
64. Helbing D. 2001 Traffic and related self-driven many-particle systems. *Rev. Mod. Phys.* **73**, 1067–1141. (doi:10.1103/RevModPhys.73.1067)
65. Ramaswamy S. 2010 The mechanics and statistics of active matter. *Annu. Rev. Condens. Matter Phys.* **1**, 323–345. (doi:10.1146/annurev-conmatphys-070909-104101)
66. Fodor É, Marchetti MC. 2018 The statistical physics of active matter: from self-catalytic colloids to living cells. *Physica A* **504**, 106–120. (doi:10.1016/j.physa.2017.12.137)
67. Onuchic JN, Wolynes PG. 2004 Theory of protein folding. *Curr. Opin. Struct. Biol.* **14**, 70–75. (doi:10.1016/j.sbi.2004.01.009)
68. Dill KA, Ozkan SB, Shell MS, Weikl TR. 2008 The protein folding problem. *Annu. Rev. Biophys.* **37**, 289–316. (doi:10.1146/annurev.biophys.37.092707.153558)
69. Wales DJ. 2003 *Energy landscapes: applications to clusters, biomolecules and glasses*. Cambridge, UK: Cambridge University Press.
70. Peshkin MA, Sanderson AC. 1988 Planning robotic manipulation strategies for workpieces that slide. *IEEE J. Robot. Autom.* **4**, 524–531. (doi:10.1109/56.20437)
71. Zumel NB. 1997 *A nonprehensile method for reliable parts orienting*. Pittsburgh, PA: Carnegie Mellon University.
72. Boothroyd G, Ho C. 1977 Natural resting aspects of parts for automatic handling. *ASME. J. Eng. Ind.* **99**(2), 314–317. (doi:10.1115/1.3439214)
73. Diedrich FJ, Warren WH. 1995 Why change gaits? Dynamics of the walk-run transition. *J. Exp. Psychol. Hum. Percept. Perform.* **21**, 183–202. (doi:10.1037/0096-1523.21.1.183)
74. Kelso JAS. 2012 Multistability and metastability: understanding dynamic coordination in the brain. *Phil. Trans. R. Soc. B* **367**, 906–918. (doi:10.1098/rstb.2011.0351)
75. Rimon E, Koditschek DE. 1992 Exact robot navigation using artificial potential functions. *IEEE Trans. Robot. Autom.* **8**, 501–518. (doi:10.1109/70.163777)
76. Full RJ, Koditschek DE. 1999 Templates and anchors: neuromechanical hypotheses of legged locomotion on land. *J. Exp. Biol.* **202**, 3325–3332.
77. Stephens GJ, Johnson-Kerner B, Bialek W, Ryu WS. 2008 Dimensionality and dynamics in the behavior of *C. elegans*. *PLoS Comput. Biol.* **4**, e1000028. (doi:10.1371/journal.pcbi.1000028)
78. Wiltschko AB, Johnson MJ, Iurilli G, Peterson RE, Katon JM, Pashkovski SL, Abraira VE, Adams RP, Datta SR. 2015 Mapping sub-second structure in mouse behavior. *Neuron* **88**, 1121–1135. (doi:10.1016/j.neuron.2015.11.031)
79. Losos JB. 1990 The evolution of form and function: morphology and locomotor performance in West Indian *Anolis* lizards. *Evolution* **44**, 1189–1203. (doi:10.1111/j.1558-5646.1990.tb05225.x)
80. Kohlsdorf Jr TG, Navas CAT. 2001 Limb and tail lengths in relation to substrate usage in Tropidurus lizards. *J. Morphol.* **248**, 151–164. (doi:10.1002/jmor.1026)
81. Domokos G, Várkonyi PL. 2008 Geometry and self-righting of turtles. *Proc. R. Soc. B* **275**, 11–17. (doi:10.1098/rspb.2007.1188)
82. Golubović A, Arsovski D, Ajtić R, Tomović L, Bonnet X. 2013 Moving in the real world: tortoises take the plunge to cross steep steps. *Biol. J. Linn. Soc.* **108**, 719–726. (doi:10.1111/bij.12000)
83. Dill KA, MacCallum JL. 2012 The protein-folding problem, 50 years on. *Science* **338**, 1042–1046. (doi:10.1126/science.1219021)
84. Voelz VA *et al*. 2012 Slow unfolded-state structuring in acyl-CoA binding protein folding revealed by simulation and experiment. *J. Am. Chem. Soc.* **134**, 12 565–12 577. (doi:10.1021/ja302528z)
85. Socci ND, Onuchic JN, Wolynes PG. 1996 Diffusive dynamics of the reaction coordinate for protein folding funnels. *J. Chem. Phys.* **104**, 5860–5868. (doi:10.1063/1.471317)
86. Byl K, Tedrake R. 2009 Metastable walking machines. *Int. J. Rob. Res.* **28**, 1040–1064. (doi:10.1177/0278364909340446)
87. Geyer H, Seyfarth A, Blickhan R. 2006 Compliant leg behaviour explains basic dynamics of walking and running. *Proc. R. Soc. B* **273**, 2861–2867. (doi:10.1098/rspb.2006.3637)
88. Tuthill JC, Azim E. 2018 Proprioception. *Curr. Biol.* **28**, R194–R203. (doi:10.1016/j.cub.2018.01.064)
89. Wilson RP, Quintana F, Hobson VJ. 2012 Construction of energy landscapes can clarify the movement and distribution of foraging animals. *Proc. R. Soc. B* **279**, 975–980. (doi:10.1098/rspb.2011.1544)
90. Anderson DJ. 2016 Circuit modules linking internal states and social behaviour in flies and mice. *Nat. Rev. Neurosci.* **17**, 692–704. (doi:10.1038/nrn.2016.125)
91. Hackett J, Gao W, Daley M, Clark J, Hubicki C. 2020 Risk-constrained motion planning for robot locomotion: formulation and running robot demonstration. In *2020 IEEE/RSJ Int. Conf. on Intelligent Robots and Systems (IROS)*, pp. 3633–3640. Las Vegas, NV: IEEE. (doi:10.1109/IROS45743.2020.9340810)
92. Suraci JP, Clinchy M, Zanette LY, Wilmers CC. 2019 Fear of humans as apex predators has landscape-scale impacts from mountain lions to mice. *Ecol. Lett.* **22**, 1578–1586. (doi:10.1111/ele.13344)
93. Berman GJ, Bialek W, Shaevitz JW. 2016 Predictability and hierarchy in Drosophila behavior. *Proc. Natl Acad. Sci. USA* **113**, 11 943–11 948. (doi:10.1073/pnas.1607601113)
94. Irschick DJ, Garland TJ. 2001 Intergrating function and ecology in studies of adaptation: investigations of locomotor capacity as a model system. *Annu. Rev. Ecol. Syst.* **32**, 367–396. (doi:10.1146/annurev.ecolsys.32.081501.114048)
95. Arnold SJ. 2003 Performance surfaces and adaptive landscapes. *Integr. Comp. Biol.* **43**, 367–375. (doi:10.1093/icb/43.3.367)
96. Burridge RR, Rizzi AA, Koditschek DE. 1999 Sequential composition of dynamically dexterous robot behaviors. *Int. J. Rob. Res.* **18**, 534–555. (doi:10.1177/02783649922066385)
97. Arslan O, Koditschek DE. 2019 Sensor-based reactive navigation in unknown convex sphere worlds. *Int. J. Rob. Res.* **38**, 196–223. (doi:10.1177/0278364918796267)
98. Tan J, Zhang T, Coumans E, Iscen A, Bai Y, Hafner D, Bohez S, Vanhoucke V. 2018 Sim-to-real: learning agile locomotion for quadruped robots. In *Robotics: Science and Systems XIV*.
99. Lee J, Hwangbo J, Wellhausen L, Koltun V, Hutter M. 2020 Learning quadrupedal locomotion over challenging terrain. *Sci. Robot.* **5986**, 1–14. (doi:10.1126/scirobotics.abc5986)
100. Kehoe B, Patil S, Abbeel P, Goldberg K. 2015 A survey of research on cloud robotics and automation. *IEEE Trans. Autom. Sci. Eng.* **12**, 398–409. (doi:10.1109/TASE.2014.2376492)

**Locomotor transitions in the potential energy landscape-dominated regime**

Ratan Othayoth, Qihan Xuan, Yaqing Wang, and Chen Li*

*Corresponding author (chen.li@jhu.edu)

**This PDF file includes:**

Table S1

Figures S1 to S8

Supplementary Text S1 to S8

Supplementary References

**Other supplementary materials for this manuscript include the following:**

Movie S1: https://youtu.be/jT7nfON-Tfc

Video abstract available at: https://youtu.be/xqFAeN9MlnQ

Overview video available at https://youtu.be/xus_Xs-Xpvw

Data, code, and interactive figures available at:

https://github.com/TerradynamicsLab/potential_energy_landscape

**Table S1. Summary of major results from reviewed studies**

| Challenges | Observed interactions | Physical principles of interaction | Implications for robotics | Category of transition strategy on energy landscape |
|---|---|---|---|---|
| Gap [2] | Animal crosses gaps as large as one body length. | High approach speed, high initial body pitch, and high initial angular velocity facilitates crossing. | Active body pitching increases traversable gap length by 50%. | Steer system state. |
| Bump [1] | Animal climbs bumps up to 4 times hip height. | High initial body pitch and low initial body yaw facilitates climbing. | Active body pitching increases traversable bump height by 75%. | Steer system state. |
| Pillar [3,4] | Cuboidal and elliptical body shapes experience obstacle attraction and repulsion, respectively. | Obstacle attraction/repulsion is an inherent property of body shape and insensitive to pillar geometry and size. | Passive control of traversal using obstacle interaction modulated by body shape<br>Elliptical body shape increased traversal probability of pillar field by 8 times<br>Active turning by legs helps cuboidal body steer away from and traverse obstacle | Modify landscape topology (shape change).<br>Steer system state (leg turning). |
| Beams [5–7] | Animal transitions between diverse locomotor modes to traverse.<br>It is more likely to transition from pitch to roll mode to traverse as beams become stiffer.<br>It flexes head, reduces leg sprawling, and uses legs differentially before rolling. | Thin, rounded body shape facilitates body rolling; body oscillation kinetic energy fluctuation facilitates barrier crossing for pitch-to-roll transitions. | Adding rounded ellipsoidal shell increases traversal probability by 4 times.<br>controlled oscillation modulates probability of locomotor transitions. | Modify landscape topology (shape change).<br>Perturb system state (kinetic energy fluctuation).<br>Steer system state (leg adjustments). |
| Self-righting [8–12] | Animal repeatedly opens wings and flails legs to self-right.<br>It rarely succeeds in somersaulting and often self-rights by rolling sideways. | Wing opening reduces barrier to roll sideways.<br>Barrier crossing is facilitated by leg flailing kinetic energy fluctuation, as well as proper wing-leg coordination, which can be achieved by randomness in actuation. | Large wing opening, asymmetric small wing opening, coordinated wing-leg oscillations facilitate self-righting.<br>Leg oscillation increases self-righting probability up to 7 times.<br>Randomness helps find proper coordination. | Change landscape barriers (wing opening).<br>Modify landscape topology (asymmetric wing opening).<br>Perturb system state (kinetic energy fluctuation). |

## Supplementary Text

### S1. Imaging tools to measure locomotor transitions and locomotor-terrain interaction

There are several technical challenges to measuring locomotor transitions and locomotor-terrain interaction during traversal of complex 3-D terrain and self-righting. First, the traditional technique of using two or three cameras for measuring 3-D motion on flat surfaces is inadequate to record the large range of 3-D rotations of the animal, robot, or terrain (if any) or cope with frequent occlusion, both common in complex 3-D terrain [5]. For example, the animal could be tracked in only 24%, 60%, or 77% of the frames during a beam traversal attempt with only 2, 3, or 4 cameras. In addition, manually measuring locomotor and terrain motion is laborious, especially for large datasets (~$10^2$–$10^3$ trials) required for making statistically meaningful conclusions when multiple locomotor modes are being observed. For example, for the ~300 animal trials in our study of beam traversal [6], it would take 500 hours to manually track the animal body and two beams. Furthermore, to measure appendage motion and frequent body-obstacle interaction smaller than body/obstacle size, we need high-accuracy 3-D kinematics. For example, a tracking error of ±1 mm in the animal's forward position during beam interaction can over or underestimate beam deflection angle by ~13°.

We developed several imaging tools to address these challenges. First, we developed an automated visible light imaging system (figure 3b) with up to 12 synchronized high-resolution, high-speed cameras [6,13]. This system can capture near-continuous locomotor-terrain interaction over the full range of 3-D translation and rotation observed and track the animal in more than 96% of the terrain interaction phase. In addition, we reduced the manual labor for calibration and 3-D motion reconstruction by 250 times by automatically tracking custom calibration objects (figure 3c) and animal and terrain motions (figure 3d) in each camera view, using uniquely distinguishable QR-code markers (figure 3a) [14]. Furthermore, to achieve high-accuracy measurement of locomotor-obstacle interaction, we designed custom calibration objects to span the entire field of view of all 12 cameras and used a high-precision 3-D printed object to verify tracking and reconstruction fidelity (s.d. of position error = 0.6 mm over a 3-D calibrated field of view of 20 cm by 20 cm; s.d. of orientation error = 1.1°).

## S2. General steps to create and use potential energy landscapes

We developed potential energy landscapes for traversal of bump [1], gap [2], pillar [3], and beam [6,7] obstacles and ground self-righting [9]. Here we summarize general steps to calculate and use them (also see figure S1).

1. Create a simplified physics model of the interaction with which system potential energy can be calculated as a function of relevant system degrees of freedom. In our studies, we focused on first understanding coarse-grained transitions between modes that differ significantly in how the body moves. Thus, to create the simplest potential energy landscape with minimal degrees of freedom, we approximated the animal or robot as a rigid body and neglected the legs. We used an ellipsoidal body shape resembling the animal body, except in the pillar study where different body shapes were studied. In addition, we assumed that the body does not penetrate the obstacle or ground and its lowest point is always in contact with the ground. See Sections S3-S7 for other assumptions specific to each model system. Finer-grained transitions between more nuanced modes that involve body bending and leg motion can, in principle, be studied using higher-dimensional potential energy landscapes, although more challenging—see reviews of energy landscape modeling of protein folding (e.g., [15]).
2. Quantify system potential energy as a function of system degrees of freedom and physical/geometrical properties. For gap, bump, pillar, and self-righting interaction where body parts are assumed rigid and terrain elements are rigid and fixed, system potential energy is the body gravitational potential energy ($E = mgz_{com}$, where $m$ is the body mass, $g$ is gravitational acceleration, and $z_{com}$ is body center of mass height). For beam interaction, in addition to body gravitational potential energy, beam gravitational energy and elastic potential energy also contribute to system potential energy.
3. Measure relevant physical/geometrical properties (such as beam stiffness, body mass/geometry) required to calculate system potential energy.

4. Choose the few system state degrees of freedom over which the potential energy landscape is to be constructed. Often degrees of freedom that represent self-propulsion (e.g., body forward position relative to obstacles, wing opening angle in ground self-righting) and those that change substantially in response to terrain interaction (e.g., body pitch, roll, yaw) are chosen. Because a high-dimensional landscape over a large number of degrees of freedom is more challenging to understand three degrees of freedom can be chosen first to visualize the landscape more easily, as a potential energy map over two degrees of freedom which further evolves over the third degree of freedom (e.g., Supplementary Movie S1). See Section S8 for visualizing the potential energy landscape over all three chosen degrees of freedom. We note that constructing the landscape over only three degrees of freedom is a compromise to simplify analysis and already provide substantial insight into experimental observations. More rigorous analysis of high-dimension landscape over all relevant degrees of freedom and comparison to experimental observations may reveal additional insight in the future.
5. Calculate the potential energy landscape over the first two chosen degrees of freedom while keeping the third (and other) degrees of freedom constant. Varying both these degrees of freedom in small increments over the desired range and calculating the potential energy at each point of the grid in this 2-D parameter space.
6. Construct the evolving potential energy landscape by repeating step 5 while varying the third (and remaining) degrees of freedom, either using an experimentally measured trajectory or prescribing a trajectory.
7. Visualize system state trajectory on the landscape, by projecting the measured or prescribed values of the first two chosen degrees of freedom on the evolving potential energy landscape. Use only the end points of the trajectory, which represent the current state, to show the actual potential energy of the system. Use the rest of the visualized trajectory to show how measurements of the first two chosen degrees of freedom evolve on the landscape. Because the potential energy landscape evolves as the third degree of freedom changes and can occlude the trajectory, project the trajectory onto the landscape surface for visualization.

8. Find local minima and identify basins of the landscape. For landscapes with simple shapes, basins can be identified from visual inspection. A more rigorous method is to use graph search algorithms [16] to first find the saddle points on the landscape and then identify the basins separated by them.
9. Determine the potential energy barrier for transitioning from one basin to another, which occurs at the saddle separating the basins. For landscapes with relatively simple shapes, this can be done in several steps. First, consider imaginary straight paths on the landscape away from one basin minimum towards the direction of another basin. Then, along each imaginary straight path, obtain a cross section of the landscape and measure the barrier along this cross section as the maximal increase in potential energy to escape from the basin along the straight path. Repeat this step to calculate potential barriers for transitioning along all possible directions to transition to the other basin. The lowest barrier among all of them is the barrier to transition from this basin to the other. See [6] for detailed steps. A more rigorous method is to calculate the potential energy increase from the basin minimum to the saddle point separating it from another basin using graph search algorithms [16].
10. Repeat steps 7-9 to measure system potential energy and potential energy barrier for transition from one basin to another as the third degree of freedom is varied.
11. Additional metrics, such as potential energy of basin minima, kinetic energy fluctuation, direction of system state velocity on the landscape, and energy landscape gradients, may be measured. Together, these help understand system behavior and principles of transitions across landscape basins.

Below, we briefly summarize specifics of calculating the potential energy landscape of each model system. See respective original studies [1–3,6,7,9] for more detail, except the landscape of gap interaction which is first presented here. Note that here we renamed some modes from those in the original studies to better distinguish modes across different model systems. See [17] for data, code and interactive plots of potential energy landscapes.

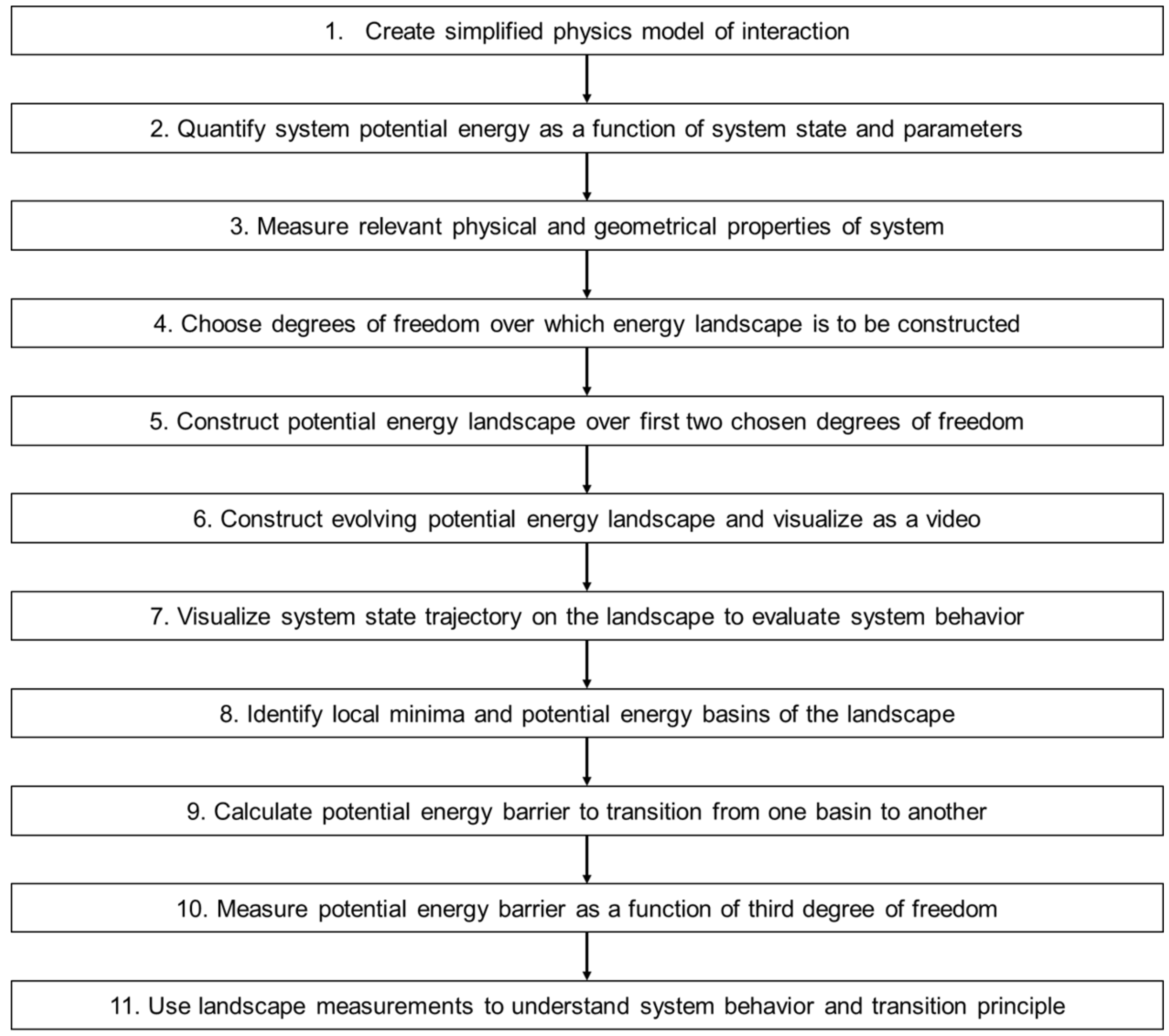

**Figure S1. General steps to calculate and use potential energy landscape of locomotor terrain interaction.**

## S3. Potential energy landscape of gap interaction

In our study of dynamic traversal of large gap obstacles, we focused on understanding and predicting successful traversal dynamics by creating a template model of gap crossing [2]. However, a potential energy landscape approach can also provide insight into the emergence of both successful crossing and falling and strategies to make the desired transition.

We calculated the potential energy landscape of body-gap interaction over the body pitch-yaw space as the body moves forward (figures S2). Throughout traversal, we constrained the body to maintain contact with the ground or the bottom of the gap and not penetrate the vertical sides of the gap. Although the body often lost contact with ground momentarily during crossing in animal and robot experiments, this constraint was required to use the potential energy landscape to model the fall mode. For a given body forward position $x$, we varied body pitch and yaw over [−90°, 90°] and calculated body potential energy from this constraint. Note that positive pitch corresponds to the body pitching head down.

Before encountering the gap, the body moves forward on level ground (figure S2a, i). In this case, potential energy depends only on body pitch and not on body yaw. The potential energy landscape has a global minimum valley along zero body pitch (figure S2c, i). As the body moves over the gap, pitching downwards lowers the center of mass and reduces potential energy (figure S2a, ii'). As a result, the initial global minimum valley shifts in the positive pitch direction (positive pitch is pitching head down), and a new fall basin develops around it (figure S2c, ii). In the fall basin, potential energy also depends on body yaw because, beyond a certain yaw, the body must pitch up to not penetrate the gap's vertical sides. Similarly, as the body continues to move forward in the gap, it must pitch down further to not penetrate the far vertical side of the gap. Alternatively, it can pitch up until its bottom contacts the farther edge of the gap. Because increasing or decreasing pitch increases potential energy, a cross basin also emerges, centered around the pitched-up state (figure S2c, iii).

Approaching the gap more slowly and/or with a lower body pitch and/or higher body yaw magnitude decreases the system's initial kinetic and/or potential energy and increases its probability of being trapped in the fall basin, resulting in the body falling into the gap (figure S2a ii'). After falling, the body can turn sideways (for a sufficiently wide gap) and move within the gap (figure S2a ii'→iii', c ii'→iii'). By contrast, approaching the gap head-on (with less body yaw), more rapidly, and/or with a higher body pitch (figure S2a ii, c ii) increases the system's initial kinetic and/or potential energy and increases its probability of overcoming the potential energy barrier to reach the cross basin (figure S2a iii, c iii), resulting in the body crossing the gap. These modeling insights are consistent with those from the dynamical template [2].

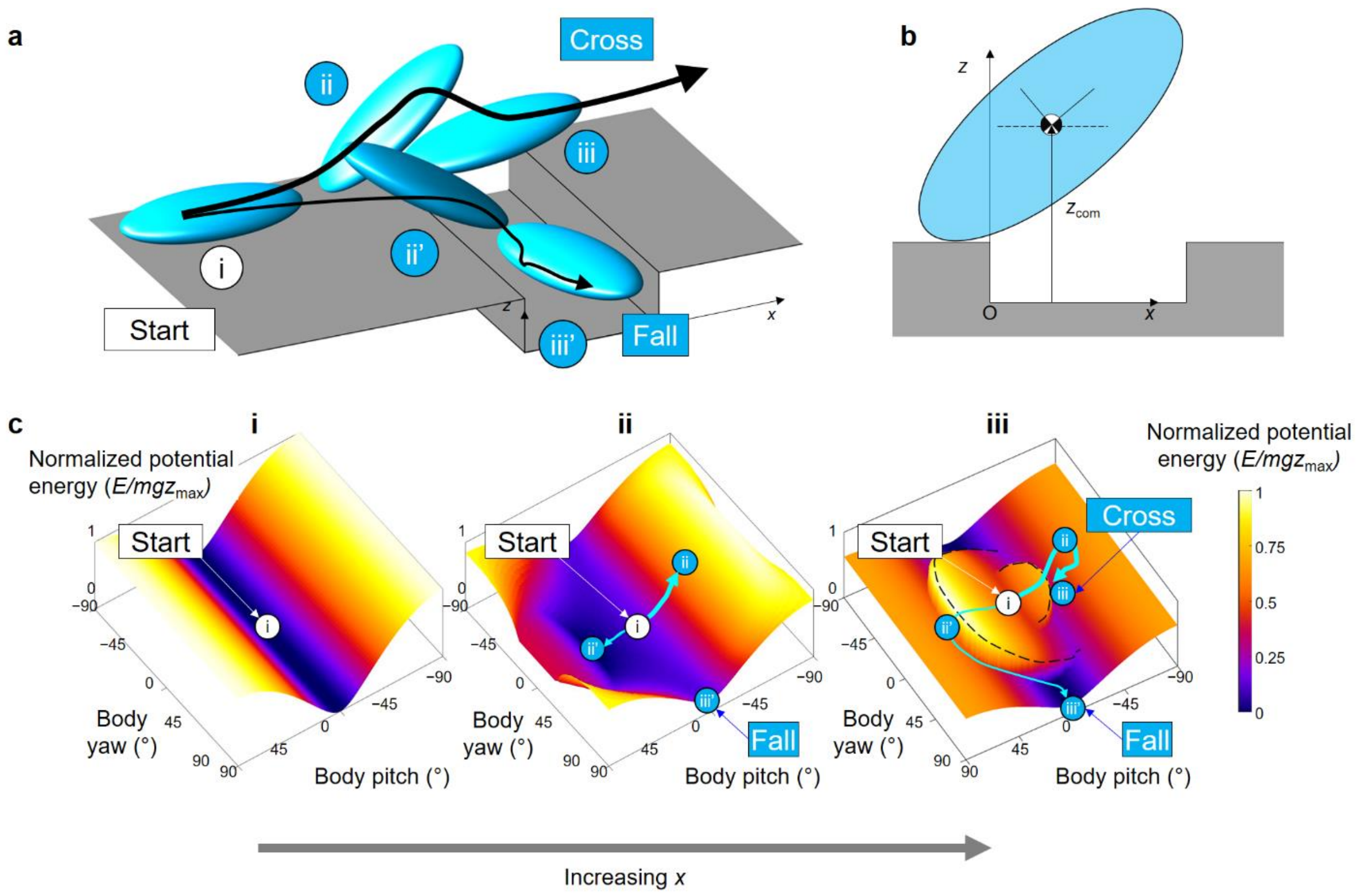

**Figure S2. Locomotor transitions on potential energy landscape of gap obstacle.** (a) Snapshots of body before (i) and during (ii and iii) interaction with gap in cross and fall modes. (b) Definition of variables and parameters. (c) Snapshots of potential energy landscape over body pitch-yaw space before (i) and during (ii and iii) interaction. Numbered dots represent corresponding system states in (a). Cyan arrows are representative state trajectories for body crossing gap (i→ii→iii) and falling into gap (i→ii'→iii'). Note that landscape evolves as body moves forward (increasing *x*). Dashed black curves show potential energy barriers separating cross and fall basins. Potential energy is normalized to maximum value possible during interaction. Note that we renamed as cross and fall modes here the traverse and fail modes in the original study to better distinguish modes across different model systems. Also see movie S1. Note from the movie that for a small range of forward position $x$, the cross basin momentarily becomes as a saddle (which is a local minimum along the pitch axis but a local maximum along the yaw axis).

## S4. Potential energy landscape of bump interaction

For a bump obstacle [1], we calculated the potential energy landscape of body-bump interaction over the body pitch-yaw space as the body moves forward (figures S3). Throughout traversal, we constrained the body to maintain contact with either the ground or top of the bump and not penetrate the ground and vertical sides of the bump. For a given body forward position *x*, we varied body pitch and yaw over [−90°, 90°] and calculated body gravitational potential energy. Note that positive pitch corresponds to the body pitching head down.

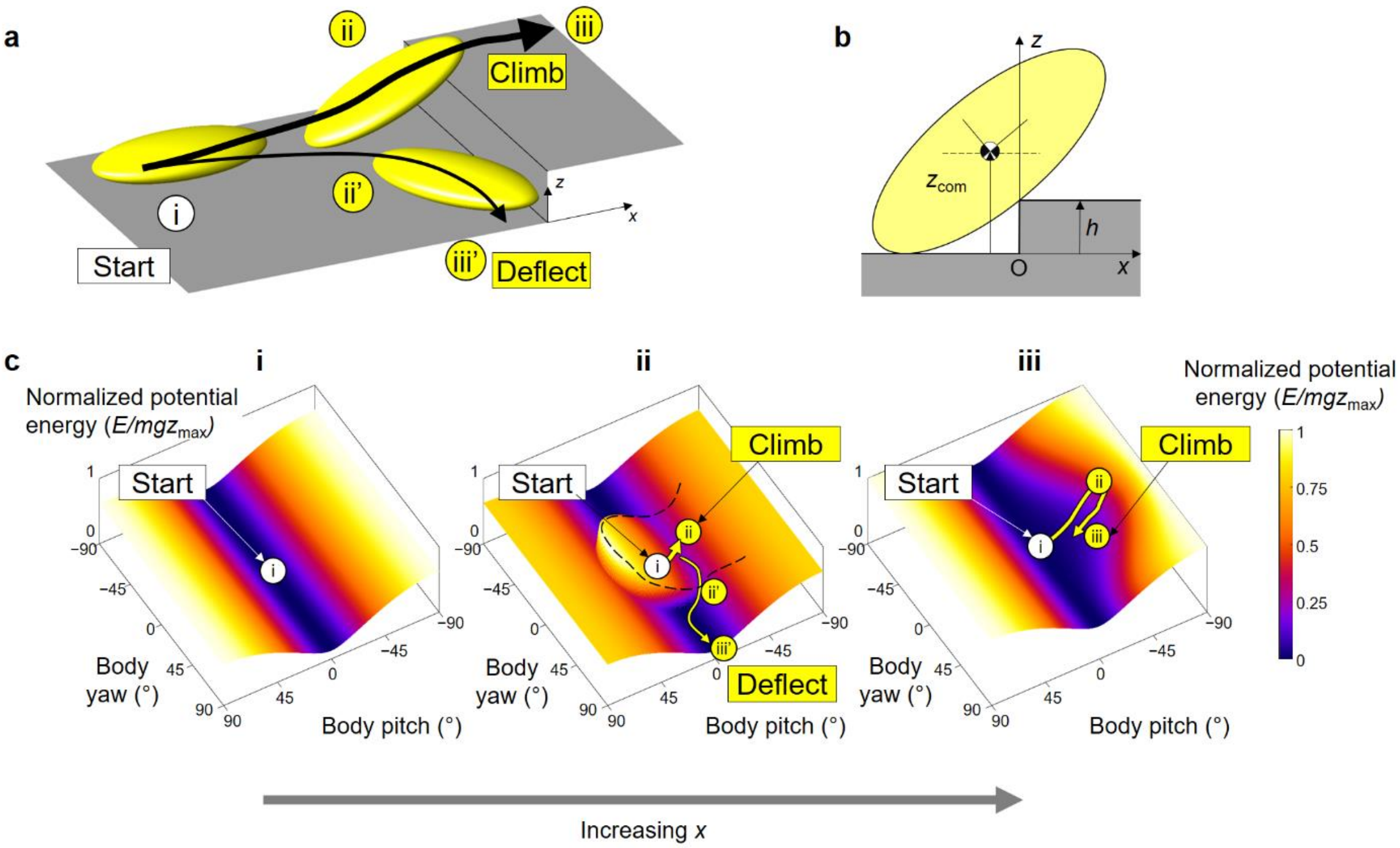

**Figure S3. Locomotor transitions on potential energy landscape of bump obstacle.** (a) Snapshots of body before (i) and during interaction with bump in climb (ii→iii) and deflect (ii'→iii') modes. (b) Definition of variables and parameters. (c) Snapshots of potential energy landscape over body pitch-yaw space before (i) and during (ii and iii) interaction. Numbered dots represent corresponding system states in (a). Yellow arrows are representative state trajectories for body climbing bump (i→ii→iii) and deflecting sideways (i→ii'→ii'). Because deflected body does not move beyond bump ($x$ = 0), deflect mode is not shown in c, iii. Note that landscape evolves as body moves forward (increasing *x*). Dashed black curves

show potential energy barriers separating climb and deflect basins. Potential energy is normalized to maximum value possible during interaction. Note that we renamed as cross and fall modes here the traverse and fail modes in the original study to better distinguish modes across different model systems. Also see movie S1. Note from the movie that for a small range of forward position $x$, the climb basin momentarily becomes a saddle (which is a local minimum along the pitch axis but a local maximum along the yaw axis).

Before encountering the bump, the body moves forward on level ground (figure S3a, i); in this case, potential energy depends only on body pitch and not on body yaw. The potential energy landscape has a global minimum valley along zero body pitch (figure S3c, i). As the body moves close to the bump, due to the constraints, it can remain horizontal only if it yaws and deflects. Thus, the initial global minimum valley splits into two deflect basins (figure S3c, ii), corresponding to the body yawing left or right. Alternatively, the body can pitch up until it does not penetrate the bump; pitching up any more or less will cause the potential energy to increase due to the constraints. As a result, a climb basin develops (figure S3c, ii). As the body continues to move forward onto the bump (figure S3c, iii), the climb basin reverts back to the initial global minimum valley, except having a higher potential energy. Note that for a small range of forward position $x$, the climb basin momentarily becomes as a saddle (which is a local minimum along the pitch axis but a local maximum along the yaw axis). In this range of $x$, the climb basin is a lower dimensional attractor with attraction only along the pitch direction (its stable manifold [18]).

See [1] for detail of animal and robot observations, landscape modeling, and strategies that facilitate climbing and suppress deflection.

**S5. Potential energy landscape of pillar interaction**

For pillar interaction, we calculated the potential energy landscape over the body pitch-bearing (turning) space as the body moved forward (figure S4). For a given body forward position ($x$, $y$), we varied body pitch and bearing over [−90°, 90°] and calculated body potential energy. Note that positive pitch corresponds to the body pitching head down. With these constraints, the potential energy depends only on body pitch and not on bearing, as the latter does not affect center of mass height.

Before encountering the pillars, the body moves forward on level ground (figure S4a, b, i). The potential energy landscape for both body shapes has a global minimum valley along zero body pitch (figure S4d, e, i). As the body moves close to the pillar, it must pitch or turn to not penetrate the pillar; the states with insufficient pitching or turning are prohibited (figure S4d, e, white regions).

As the cuboidal body moves closer to the pillar, two prohibited regions emerge on both sides of a climb basin (figure S4d, ii). As the body continues to move forward, the local minimum of climb basin, surrounded by the two prohibited regions, shifts towards higher body pitch (figure S4d, ii→iii). The body can traverse the pillar by pitching and turning to overcome potential energy barriers and escaping the climb basin (figure S4d, ii→iii'→iv'). If the body pitches upward by more than 90°, it flips backwards (figure S4d, ii→iii→iv).

As the elliptical body moves close to the pillar, a single prohibited region emerges at the center of the landscape (figure S4e, i→ii). As the body moves forward, this prohibited region becomes larger. Instead of pitching and climbing, the elliptical body can traverse the pillar by turning sideways to retain a horizontal posture by escaping to the turn basin (figure S4e, ii→iii) and continue moving forward (figure S4e, iii→iv).

See [3] for detail of animal and robot observations, landscape modeling, and strategies that facilitate traversal, suppress climbing, and transition from climbing to traversal.

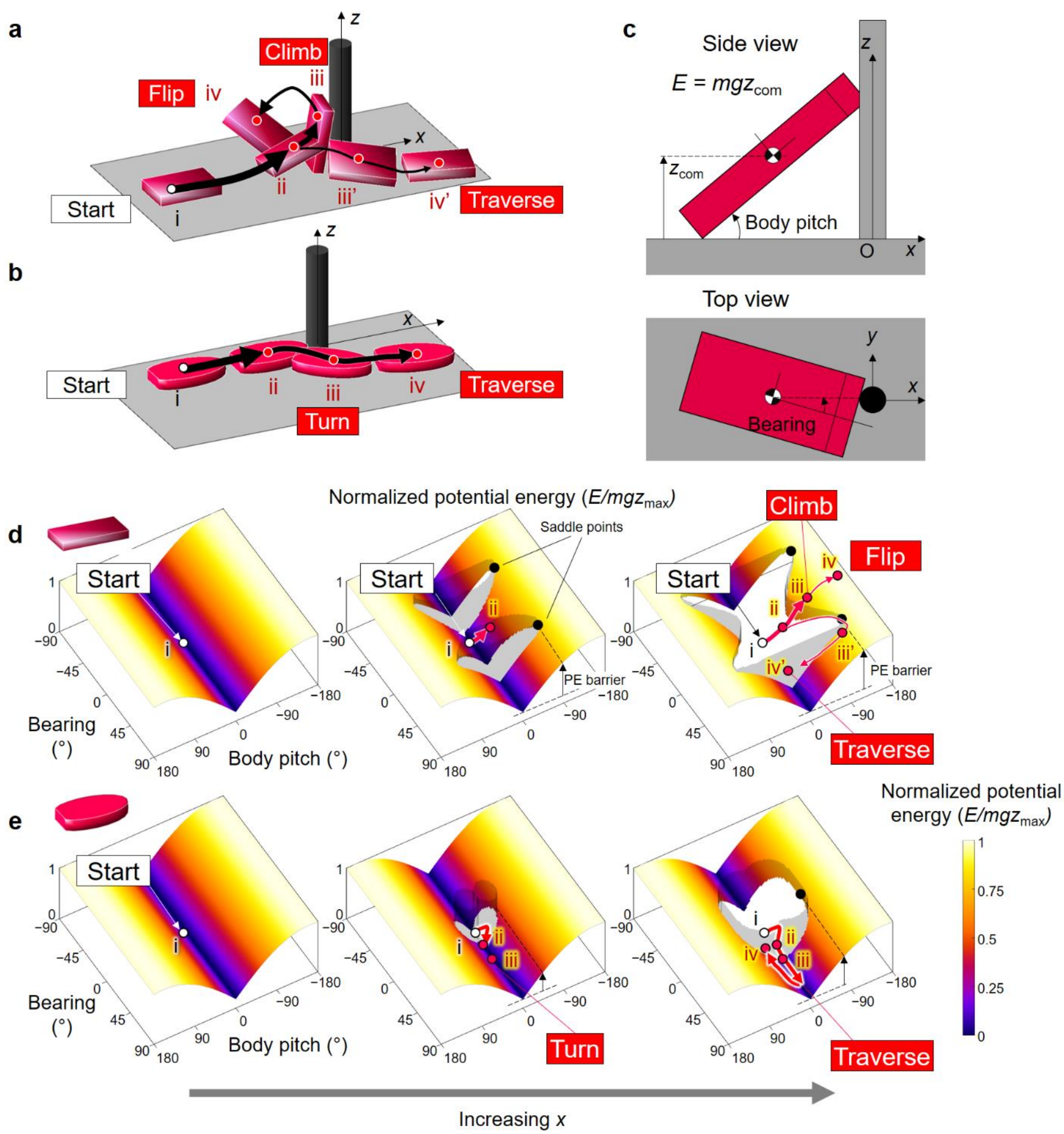

**Figure S4. Locomotor transitions on potential energy landscape of pillar obstacle.** (a, b) Snapshots of body before (i) and during (ii-iv/iv') interaction with the pillar for cuboidal (a) and elliptical (b) body shapes. (c) Definition of variables and parameters, using cuboidal body shape as example. (d) Snapshots of potential energy landscape over body pitch-bearing space before (left) and during (center and right) interaction for cuboidal body shape. Numbered dots represent corresponding system states in (a). Red arrows are representative state trajectories for cuboidal body first pitching up against pillar (i→ii),

continuing to climb against pillar (ii→iii), and finally flipping over (iii→iv), or turning sideways and traversing (ii→iii'→iv'). (e) Snapshots of potential energy landscape over body pitch-bearing space before (left) and during (center and right) interaction for elliptical body shape. Representative state trajectories for elliptical body first contacting pillar (i→ii) and then turning away from pillar and traversing (ii→iii→iv). Note that landscape evolves as body moves forward (increasing $x$). Potential energy is normalized to maximum value possible during interaction. Note that we renamed as climb mode here the turn towards and pitch up mode in the original study [3] to better distinguish modes across different model systems. Also see movie S1.

## S6. Potential energy landscape of beam interaction

For beam interaction, we calculated the potential energy landscape over the body pitch-roll space as the body moves forward (figure S5). We modeled the beams as rigid rectangular plates on torsional joints and the torsional joint as a perfect Hookean torsional spring. Because the animal or robot almost always pushed forward against the beams, in the model we only allowed forward beam deflection. Forward deflection lowers the beam center of mass and thus beam gravitational potential energy. We calculated system potential energy $E$ by summing up body and beam gravitational potential energy and beam elastic potential energy:

$$E = m_{\text{body}} g \Delta z + \frac{1}{2} m_{\text{beam}} g L(\cos\Delta\theta_1 + \cos\Delta\theta_2 - 2) + \frac{1}{2} K(\Delta\theta_1^2 + \Delta\theta_2^2) \quad (1)$$

where $m_{\text{body}}$ is body mass, $g$ is gravitational acceleration, $\Delta z$ is body center of mass height increase from its equilibrium configuration (at zero pitch and zero roll) without beam contact, $m_{\text{beam}}$ is beam mass, $L$ is beam length, $K$ is beam torsional stiffness, and $\Delta\theta_1$ and $\Delta\theta_2$ are beam deflection angles from vertical, with $\Delta\theta_{1,2} \geq 0$ (figure S5b). For a given body horizontal position ($x$, $y$) and body yaw, we varied body pitch and roll over [−180°, 180°] and calculated system potential energy (only a smaller relevant range of body pitch and roll is shown in figures S5).

Before encountering the beams, the body moves forward on level ground (figure S5a, i), and the potential energy landscape has a global minimum at zero body pitch and zero body roll (figure S5c, i). As the body moves closer to the beams, the beams are deflected forward for certain body orientations, and the global minimum evolves into a pitch basin around a local minimum at a finite body pitch and zero body roll (figure S5c, ii). This corresponds to the body pitching up and pushing against the beams (figure S5a, ii, iii). In addition, two roll basins emerge on both sides of the pitch basin (figure S5c, ii, iii), corresponding to the body rolling left or right into the gap between the two beams with minimal beam deflection. To traverse, the initially pitched-up body (figure S5a, c, i→ii) can either continue to push against the beams while maintaining a positive pitch (figure S5c, ii→iii) or roll into the gap (figure S5c, ii→iii').

See [6] for detail of animal and robot observations, landscape modeling, and see [6,7] for strategies that facilitate transition from pitching to rolling.

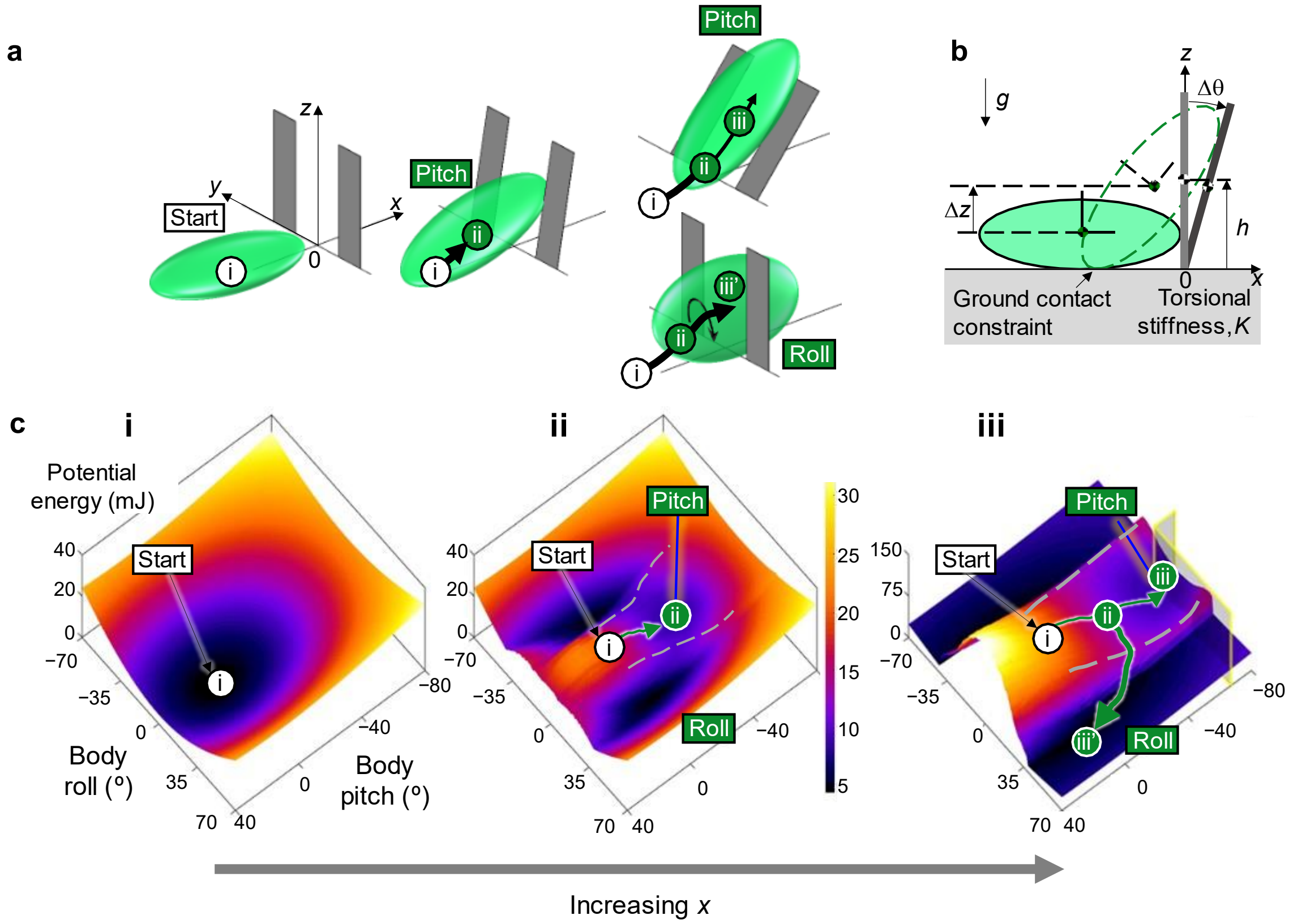

**Figure S5. Locomotor transitions on potential energy landscape of beam obstacle.** (a) Snapshots of body before (i) and during (ii and iii) interaction with the beams in pitch and roll mode. (b) Definition of variables and parameters. (c) Snapshots of potential energy landscape over body pitch-roll space before (i) and during (ii and iii) interaction. Numbered dots represent corresponding system states in (a). Green arrows are representative state trajectories for body pitching up (i→ii) and pushing across the beams (ii→iii) or by rolling into beam gap to traverse (ii→iii'). Note that landscape evolves as body moves forward (increasing *x*). Dashed grey curves show potential energy barriers separating pitch and roll basins. Also see movie S1.

## S7. Potential energy landscape of self-righting

For self-righting, we calculated the potential energy landscape over the body pitch-roll space as the wings open (figure S6). During self-righting, center of mass height and thus potential energy depend not only on body shape [8] but also on wing opening [12] and leg flailing [9]. However, given the difficulties in measuring and quantifying the animal's highly variable motion [8,9], we chose to focus potential energy landscape modeling on the simplified robotic physical model. We approximated all parts of the robot to be rigid and of uniform density.

Because the effect of leg flailing perturbation was modelled using kinetic energy fluctuation, we set the robot's leg to be held fixed in the middle of the body when calculating the potential energy landscape. We varied wing opening angle within [0º, 90°]. For each wing opening angle, we then varied body pitch and roll within [−180°, 180°] and calculated the system potential energy for each combination of body pitch and body roll (figure S6). Initially, with the wings closed, the body is horizontal and upside-down (figure S6a, i), and the system is in an upside-down basin (figure S6c, i). As the wings open, the body pitches forward. It enters a metastable state, with the center of mass projecting down into a triangular support on the ground formed by the head and both wings (figure S6a, i→ii). On the landscape, this process corresponds to the upside-down basin evolving into a metastable basin around a local minimum with a positive pitch and zero roll (figure S6c, i→ii). To self-right, the body can either continue to pitch to complete a somersault (figure S6a, c, ii→iii→iv) or roll sideways (figure S6a, c, ii→iii'→iv') to reach one of the upright basins.

See [9] for detail of animal and robot observations, landscape modeling, and see [8–12] for strategies that facilitate transition from pitching to rolling.

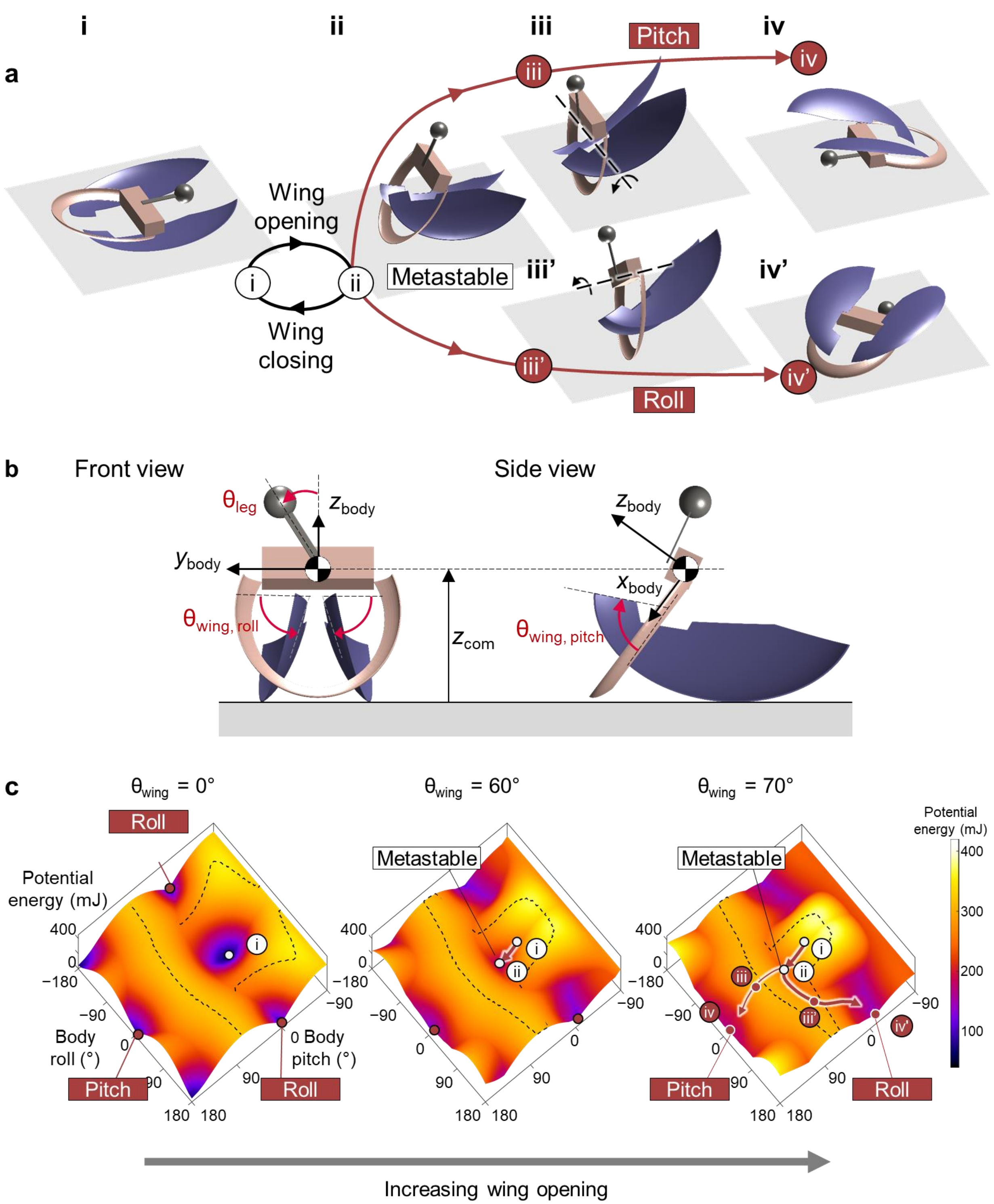

**Figure S6. Locomotor transitions on potential energy landscape of self-righting.** (a) Snapshots of robot during its wing opening (i→ii) and closing (ii→i) attempts to self-right by pitching and somersaulting (ii→iii→iv) and by rolling sideways (ii→iii’→iv’). (b) Definition of variables. (c) Snapshots of potential

energy landscape over body pitch-roll space with different wing opening. Numbered dots represent corresponding system states in (a). Brown arrows show representative state trajectories for body first pitching up to reach metastable state (i→ii), continuing to pitch and somersault (ii→iii→iv), or rolling sideways (ii→iii'→iv'). Note that landscape evolves as wings open or close (changing $\theta_{wing}$). Dashed black curves show potential energy barriers separating upside-down/metastable from pitch upright and roll upright basins. Also see movie S1.

## S8. Visualizing higher dimensional potential energy landscape

Because we consider system potential energy in the space of three chosen degrees of freedom, the energy landscape is a 4-D surface, with potential energy as the fourth dimension. For each model system, we created an interactive figure to show the cross section of the 4-D surface over each of the three chosen degrees of freedom [17]. Note that for pillar interaction landscape, no color is shown for the prohibited regions in the 3-D parameter space. Such analysis is useful—for example, we used the potential energy landscape to calculate the minimal torque required to steer the system state from pitch to roll basin during beam traversal and demonstrated the controlled transition using sensory feedback [19].

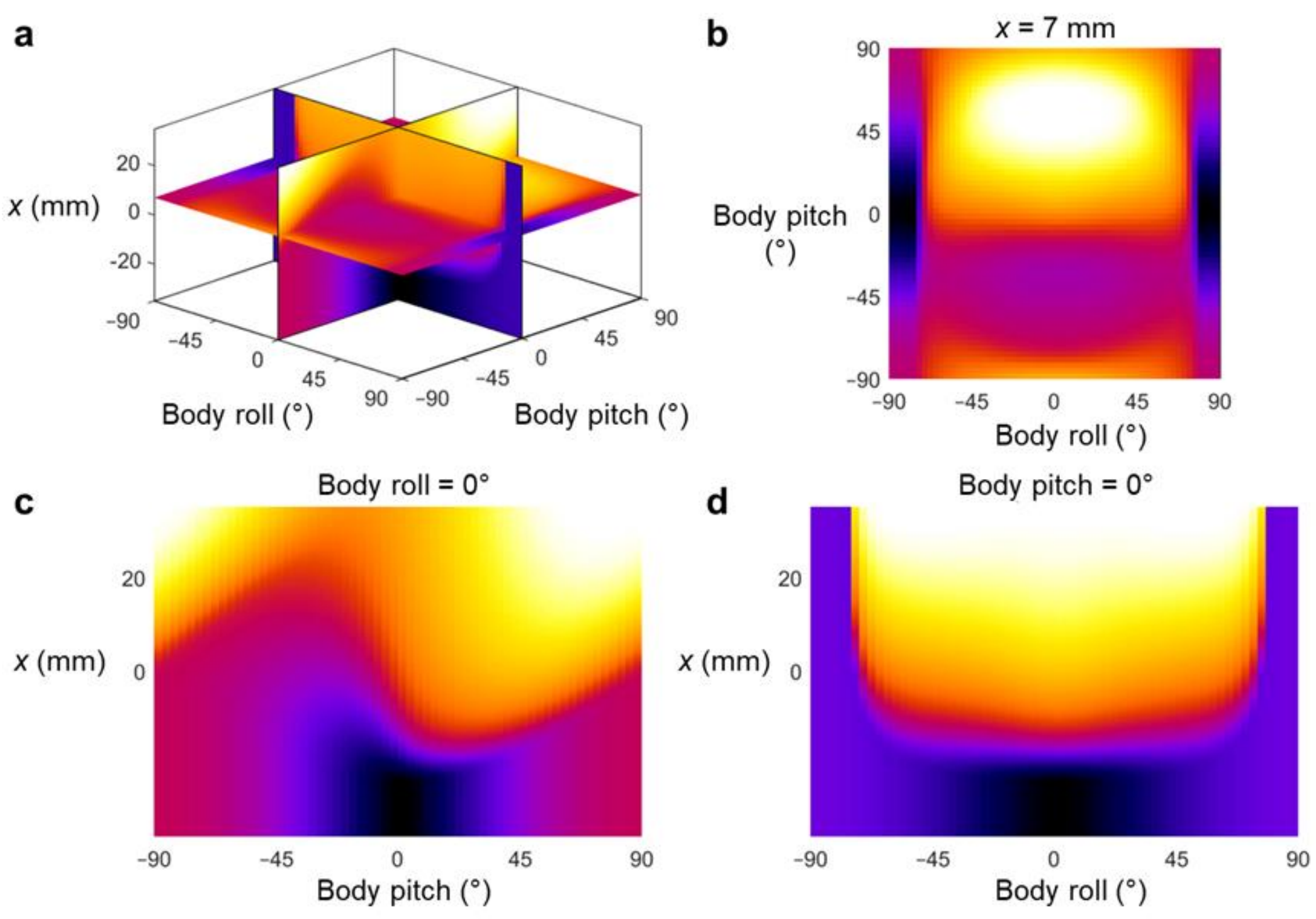

**Figure S7. Visualization of potential energy landscape over three chosen degrees of freedom.** Shown for beam interaction as an example. (a) Potential energy landscape over the three chosen degrees of freedom (body pitch, roll, and forward position ($x$) for beam interaction). (b, c, d) Cross sections of landscape orthogonal to each chosen degree of freedom. See interactive visualizations and data of each model system at [17].

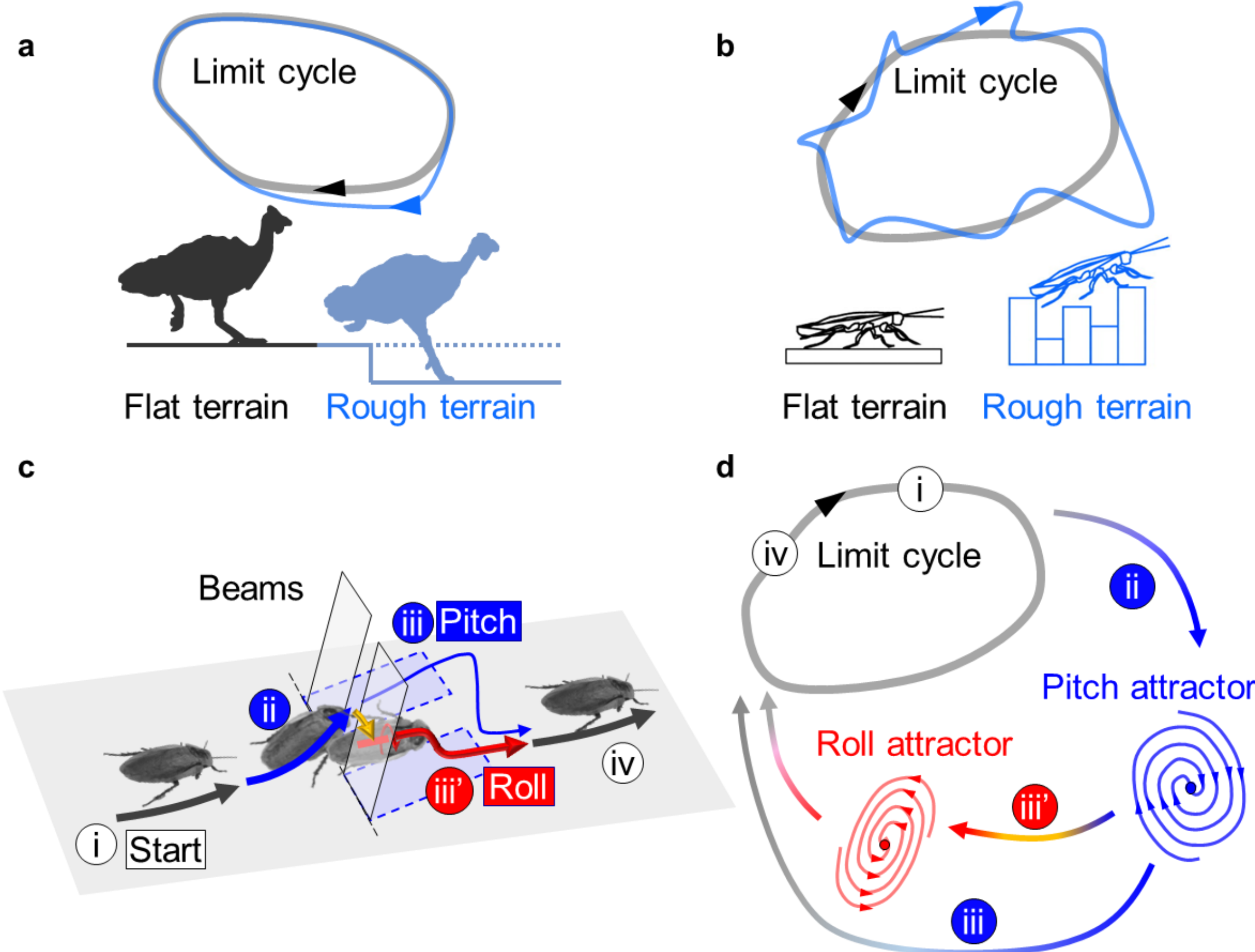

**Figure S8. Envisioned dynamical systems theories of locomotor transitions in the potential energy landscape-dominated regime.** (a) During stable walking and running on flat ground, an animal or a robot is transiently destabilized from an attractive limit cycle by occasional small obstacles [20] but quickly return to it. (b) In rough terrain, the animal or robot is continually perturbed substantially, whose dynamics is not well described by a limit cycle. (c, d) We posit that animals and robots must destabilize from limit cycles to transition between different landscape basin attractors when moving in the potential energy landscape-dominated regime. Spiral sinks [18] are used as a speculative schematic for landscape basin attractors; their exact nature remains to be discovered.

**Supplementary References**

1. Gart SW, Li C. 2018 Body-terrain interaction affects large bump traversal of insects and legged robots. *Bioinspir. Biomim.* **13**, 026005. (doi:10.1088/1748-3190/aaa2d0)
2. Gart SW, Yan C, Othayoth R, Ren Z, Li C. 2018 Dynamic traversal of large gaps by insects and legged robots reveals a template. *Bioinspir. Biomim.* **13**, 026006. (doi:10.1088/1748-3190/aaa2cd)
3. Han Y, Othayoth R, Wang Y, Hsu C-C, de la Tijera Obert R, Francois E, Li C. 2021 Shape-induced obstacle attraction and repulsion during dynamic locomotion. *Int. J. Rob. Res.* , 027836492198937. (doi:10.1177/0278364921989372)
4. Othayoth R, Francois E, Li C. 2021 Large spatiotemporal scale measurement of cockroach traversal of large obstacles. In *Integrative and Comparative Biology 61*, pp. e673–e674. (doi:10.1093/icb/icab001)
5. Li C, Pullin AO, Haldane DW, Lam HK, Fearing RS, Full RJ. 2015 Terradynamically streamlined shapes in animals and robots enhance traversability through densely cluttered terrain. *Bioinspir. Biomim.* **10**, 46003. (doi:10.1088/1748-3190/10/4/046003)
6. Othayoth R, Thoms G, Li C. 2020 An energy landscape approach to locomotor transitions in complex 3D terrain. *Proc. Natl. Acad. Sci.* **117**, 14987–14995. (doi:10.1073/pnas.1918297117)
7. Wang Y, Othayoth R, Li C. 2021 Uncovering the role of head flexion during beam obstacle traversal of cockroaches. In *Integrative and Comparative Biology 61*, p. e959. (doi:10.1093/icb/icab001)
8. Li C, Wöhrl T, Lam HK, Full RJ. 2019 Cockroaches use diverse strategies to self-right on the ground. *J. Exp. Biol.* **222**, jeb186080. (doi:10.1242/jeb.186080)
9. Othayoth R, Xuan Q, Li C. 2021 Co-opting propelling and perturbing appendages facilitates strenuous ground self-righting. *bioRxiv* (doi:https://doi.org/10.1101/2021.04.06.438657)
10. Xuan Q, Li C. 2020 Coordinated appendages accumulate more energy to self-right on the ground. *IEEE Robot. Autom. Lett.* **5**, 6137–6144. (doi:10.1109/LRA.2020.3011389)
11. Xuan Q, Li C. 2020 Randomness in appendage coordination facilitates strenuous ground self-

righting. *Bioinspir. Biomim.* **21**, 1–9. (doi:10.1088/1748-3190/abac47)

12. Li C, Kessens CC, Fearing RS, Full RJ. 2017 Mechanical principles of dynamic terrestrial self-righting using wings. *Adv. Robot.* **31**, 881–900. (doi:10.1080/01691864.2017.1372213)

13. Fu Q, Li C. 2020 Robotic modelling of snake traversing large, smooth obstacles reveals stability benefits of body compliance. *R. Soc. Open Sci.* **7**. (doi:10.1098/rsos.191192)

14. Crall JD, Gravish N, Mountcastle AM, Combes SA. 2015 BEEtag: A low-cost, image-based tracking system for the study of animal behavior and locomotion. *PLoS One* **10**, 1–13. (doi:10.1371/journal.pone.0136487)

15. Wales DJ. 2003 *Energy Landscapes: Applications to Clusters, Biomolecules and Glasses*. Cambridge University Press.

16. Cormen TH, Leiserson CE, Rivest RL, Stein C. 2009 *Introduction to Algorithms, Third Edition*. The MIT Press. (doi:10.5555/1614191)

17. Othayoth R, Xuan Q, Wang Y, Li C. 2021 Data, code, and figures accompanying publication: 10.1098/rspb.2020.2734. GitHub. See https://github.com/TerradynamicsLab/potential_energy_landscape.

18. Strogatz S. 2014 *Nonlinear Dynamics and Chaos: With Applications to Physics, Biology, Chemistry, and Engineering*. 2nd edn. Avalon Publishing.

19. Xuan Q, Li C. 2021 An energy landscape based dynamic model to simulate locomotion in complex 3-D terrain. *Integr. Comp. Biol.* **61**, e1000. (doi:10.1093/icb/icab001)

20. Daley MA, Biewener AA. 2006 Running over rough terrain reveals limb control for intrinsic stability. *Proc. Natl. Acad. Sci.* **103**, 15681–15686. (doi:10.1073/pnas.0601473103)